\documentclass[aps,pre,twocolumn]{revtex4-2}
	\usepackage{amsmath}
	\usepackage{graphicx}
	\usepackage{dcolumn}
	\usepackage{bm}
	\usepackage{gensymb}
	\usepackage{amsfonts}

\begin{document}
	
	\title{Stochastic Reconstruction of the Speed of Sound in Breast Ultrasound Computed Tomography with Phase Encoding in the Frequency Domain}
	\author{Luca A. Forte}
	 \email{lforte001@dundee.ac.uk}
	\affiliation{School of Science and Engineering, University of Dundee, Nethergate, Dundee, DD1 4HN, Scotland, UK}
	
	\date{\today}
	
	\begin{abstract}
	The framework of ultrasound computed tomography (USCT) has recently re-emerged as a powerful, safe and operator-independent way to image the breast. State of the art image reconstruction methods are performed with iterative techniques based on deterministic optimization algorithms in the frequency domain in the 300 kHz - 1 MHz bandwidth. Alternative algorithms with deterministic and stochastic optimization have been considered in the time-domain. In this paper, we present the equivalent stochastic inversion in the frequency domain (phase encoding), with a focus on reconstructing the speed of sound. We test the inversion algorithm on synthetic data in 2D and 3D, by explicitly differentiating between inverse crime and non-inverse crime scenarios, and compare against the deterministic inversion. We then show the results of the stochastic inversion in the frequency domain on experimental data. By leveraging on the concepts of multiple super-shots and stochastic ensembles, we provide robust evidence that image quality of a stochastic reconstruction of the speed of sound with phase encoding in the frequency domain is comparable, and essentially equivalent, to the one of a deterministic reconstruction, with the further benefit of drastically reducing reconstruction times by more than half.
	\end{abstract}
	
		\maketitle
		
		\section{Introduction}
		
		The birth of ultrasound computed tomography can be traced back to only a few years after the X-ray CT revolution, mostly in the context of breast imaging. In particular, ultrasound transmission tomography was pioneered in \cite{Greenleaf}; tomographic geometries in reflection mode were considered in \cite{Norton1} (circular ring) and \cite{Norton2}  (spherical bowl). A system employing both reflection data and transmission data was described in \cite{Carson}; in the latter, the authors built three sources of contrast, a reflectivity image enabled by pulse echo data, 
a speed of sound image and an attenuation image enabled by transmission data. Further investigations of USCT were considered in \cite{Kurahashi} and \cite{Honda}, with a focus on diffraction tomography algorithms. A system with a multi-frequency excitation and diffraction tomography in the frequency domain was considered in \cite{Andre}. The advent of iterative algorithms led to a major breakthrough, with the publication of the patent \cite{Johnson}. The authors presented an iterative algorithm to reconstruct the acoustic properties of human body based on the solution of the full wave equation in the frequency domain (full waveform inversion, FWI); these methods had initially been developed within the seismic imaging community and subsequently translated to medical imaging.

USCT has recently reached a level of (clinical) maturity
in the 2D (\cite{DuricSD}-\cite{DuricPMB}) and 3D (\cite{Wiskin1}-\cite{Wiskin2}) systems commercialized by Delphinus Medical Technologies and QT Imaging respectively. These systems have been cleared as adjoint imaging tools for screening mammography and their performance is currently being investigated. Both systems are capable of producing high quality images (reflectivity, speed of sound and attenuation) that can match image quality of an MRI scan (the latter often with contrast injection). Following propagation of a wide-band pulse, both systems record reflection, diffraction and transmission data via a conventional RF direct sampling scheme. Both systems have been designed to work at frequencies higher than 300-400 kHz. Image reconstruction is performed iteratively, first in the time-domain (travel-time tomography, \cite{DuricTT}, \cite{Vetterli1}) and then in the frequency-domain (FD-FWI, \cite{DuricPMB}), the latter in the 300 kHz - 1 MHz bandwidth \footnote{For completeness, we have to mention that non-iterative imaging algorithms based on a combination of beamforming and diffraction tomography techniques have also been tested on the circular array commercialized by Delphinus, images are shown in \cite{Simonetti1}-\cite{Simonetti2}. These, however, don't seem to provide the same image quality as the one obtained with FD-FWI.}.

Alternative algorithms exploiting iterative methods in the time-domain are indeed available. These solve the full wave equation in the time-domain (TD-FWI); they are described for example in \cite{Vetterli2} and in \cite{HuangIP}. Because of memory requirements and much longer computing times, TD-FWI is not a a viable option for medical imaging; this explains the choice of the previous systems to employ FD-FWI. To overcome the limitations of conventional TD-FWI, the seismic imaging community has developed \textit{stochastic} inversion techniques generally known as source encoding \cite{Krebs}. These have been translated to USCT in \cite{Huang}, which, although in a succinct formulation, considers a very general source encoding scheme. A specialized case (polarity) was considered in \cite{Anastasio1} and \cite{Anastasio2}: in these two papers the authors present results both for synthetic data and experimental data acquired with the Delphinus system (experimental phantom and patients data). Another specialized case (random time delays) was considered in \cite{Lucka}, where results on synthetic data for a 3D imaging bowl were presented. A variation of the source encoding scheme was discussed in \cite{Tromp}, on synthetic data and experimental phantom data; in this paper, the authors have extended the stochastic reconstruction to attenuation maps too.

In this paper, we present the equivalent stochastic iterative inversion in the frequency domain. The paper is organized as follows. In Section \ref{fd_fwi}, we review conventional full waveform inversion in the frequency domain and discuss the stochastic formulation. In Section \ref{Results}, we extensively test both reconstruction algorithms on synthetic data and publicly available experimental data. Finally, in Section \ref{sec:end}, we summarize our findings and we provide a critical discussion of some aspects of the presented methods and related questions.
	
	\section{Waveform inversion in the Frequency-Domain} \label{fd_fwi}
	Propagation of sound waves in the frequency-domain is accurately described by the Helmholtz equation:
	\begin{equation}
	\left(\nabla^{2} + \frac{\omega^{2}}{c^{2}(\boldsymbol{x})}\right) \, p(\omega,\boldsymbol{x}; c) = -s(\omega,\boldsymbol{x})
	\end{equation}
	where $\nabla^{2}$ is the standard Laplacian in Cartesian coordinates, $\omega = 2\pi f$, with $f$ the physical (linear) frequency measured in Hz, and $c(\boldsymbol{x})$ the (generally) spatially varying speed of sound; for simplicity, we ignore inhomogeneties in the mass density. The right hand side of the equation corresponds to a source emitting a continuous signal at a fixed frequency $\omega$, placed at Cartesian coordinate $\boldsymbol{x}$. The previous equation can  be written as
	\begin{equation}
	A\left(\omega, \boldsymbol{x};\,c(\boldsymbol{x})\right) p\left(\omega,\boldsymbol{x}; c(\boldsymbol{x})\right) = -s(\omega,\boldsymbol{x})
	\end{equation}
	where the differential operator $A$ is the imaging operator in the frequency domain. Wave propagation, i.e. the acosutic pressure, is linear with respect to the source term: if $p_{i_{1}}$ is the acoustic pressure driven by source $s_{i_{1}}$ and $p_{i_{2}}$ is the acoustic pressure driven by source $s_{i_{2}}$ (with $s_{i_{1}}$ and $s_{i_{2}}$ distinct in space), then the pressure $p_{i_{1},\,i_{2}}$ driven by having both sources on is trivially the sum, i.e. $p_{i_{1},\,i_{2}} =p_{i_{1}} + p_{i_{2}}$ with $A\, p_{i_{1},\,i_{2}} =-(s_{i_{1}} + s_{i_{2}})$. Note, however, that the the imaging operator depends non-linearly on the speed of sound: the function $f: c \rightarrow p(c)$ that maps a known speed of sound $c$ to the acoustic pressure $p(c)$ is non-linear. The Helmholtz equation allows then to calculate the acoustic pressure at all points in space for a fixed frequency. As it is the case for any PDE in a continuous setting, the Helmholtz equation has to be complemented with boundary conditions. In our specific case of a closed geometry sensing an object with the acoustic pressure measured at a discrete number of sensors (continuous to discrete imaging model), the appropriate boundary conditions are given by the well known Sommerfield radiation conditions, which dictate precise decaying rules for the pressure at infinity (the reader interested in the mathematical aspects of acoustic inverse scattering in a continuous setting is referred to the beautiful book \cite{ColtonKress}). In a discrete setting, these boundary conditions are usually implements via absorbing boundary conditions or perfectly matched layers. In the following description of our discrete-to-discrete imaging model, all vector or matrix quantities are denoted in bold. In medical imaging or similar applications like non-destructive testing, one is interested in recovering the speed of sound from the measured sensors data, $g: \boldsymbol{d}^{(obs)} \rightarrow \boldsymbol{c}$. This is a prototype of a so-called inverse problem (acoustic tomography). We indicate the estimate of the speed of sound by the same letter $\boldsymbol{c}$ with an abuse of notation. It's worth stressing that the observed data $\boldsymbol{d}^{(obs)}$ have an implicit dependence on the  source term $s$. In an experimental setting, the source term is controlled by transmitting a known voltage waveform, hence we can safely assume that the mathematical representation of the source term is known both in the forward model and in the inverse problem, and we can drop the explicit dependence of the observed data on the excitation waveform. In the general case, there are no known analytical expressions for the speed of sound. Thanks to technological advances in computing (e.g.  GPU), it is now possible to invert for the speed of sound by employing iterative techniques based on a physical model (model-based image reconstruction).	The most accurate inversion algorithm solves the full wave equation and it is therefore known as full waveform inversion. This is an example of a more general class of model-based image reconstruction methods known as PDE-constrained optimization. These can be written as
			\begin{equation}
		\begin{cases}
       \hat{ \boldsymbol{\theta }} = \arg\min_{\boldsymbol{{\theta}}} \left\Vert  \boldsymbol{d}^{(syn)}(\boldsymbol{\theta}) - \boldsymbol{d}^{(obs)}  \right\Vert^{2}_{2}\\
       \boldsymbol{M}(\boldsymbol{\theta}) =  0
    \end{cases}       
	\end{equation}
	where the quantity $\boldsymbol{\theta}$ represents one or more parameters of the model (in the present case, $\boldsymbol{\theta} = \boldsymbol{c}$) and the constraint $\boldsymbol{M}$ (i.e. the physics) is described by a PDE (in the present case, the Helmholtz equation). FWI is an instance of a non-linear least squares method aiming to minimize a misfit functional between model-predicted synthetic data and observed data. Although FWI can also be formulated in the time-domain, in this paper we focus on FD-FWI; in this case, if $N_{f}$ is the number of frequencies, the cost function can be written as
		\begin{equation} \label{CostFunctionFF}
	C(\boldsymbol{c}) = \sum_{\omega_{i}\,=\,1}^{N_{f}} C(\omega_{i};\boldsymbol{c})  = \sum_{\omega_{i}\,=\,1}^{N_{f}} \left\Vert\boldsymbol{d}^{(syn)}({\omega_{i}};\boldsymbol{c})  - \boldsymbol{d}^{(obs)}({\omega_{i}})\right\Vert_{2}^{2}
	\end{equation}
where the nature of synthetic data and observed data as frequency samples has been made explicit.
	\subsection{Single-Frequency Deterministic Optimization}
	Any iterative inversion aimed to minimize a cost function requires an initial guess (in the current case, an initial velocity model) and the calculation of the numerical gradient. By linearity, the gradient of the previous cost function, equation (\ref{CostFunctionFF}), is equal to the sum of the single-frequency numerical gradients. The calculation of the single-frequency gradient is discussed below. This approach to FD-FWI was originally formulated in \cite{Pratt}; good reviews may be found for example in \cite{Plessix} and in \cite{Norton}.
	The iterative updates of the speed of sound are written as $\boldsymbol{c}_{(k)}(\boldsymbol{x})$, where $\boldsymbol{c}_{(0)}(\boldsymbol{x})$ is a known initial guess. 
	Any model-based image reconstruction algorithm requires a numerical mathematical modeling of the underlying physics: in this case, the imaging physics is all contained in the acoustic impedance matrix $\boldsymbol{A}$. This can be built by standard finite difference methods for the spatial derivatives. Without loss of generality, we consider the case of a sensing device with $N_{Tx}$ transmitting elements and $N_{Rx}$ receiving elements; these elements may be physically distinct. In a real device, the transmitting elements fire sequentially and the acoustic pressure is usually measured at all available receiving elements. For simplicity, in the following we omit the dependence of the speed of sound on the spatial coordinated $\boldsymbol{x}$; we also fix the value of the frequency $\omega = \omega_{0}$. The first step is the calculation of the pressure at all receiving elements assuming a current velocity model $\boldsymbol{c}_{(k)}$:
	\begin{equation} \label{eq:ForwardD}
	\boldsymbol{A}\left(\omega_{0}, \boldsymbol{x};\,\boldsymbol{c}_{(k)}\right)\, \boldsymbol{p}_{i_{Tx}}\left(\omega_{0},\boldsymbol{x}; \boldsymbol{c}_{(k)}\right)\,= -\boldsymbol{s}_{i_{Tx}}\left(\omega_{0},\delta(\boldsymbol{x} - \boldsymbol{x}_{i_{Tx}})\right)
	\end{equation}
		The source term is a complex variable as large as the entire computational grid, in particular the quantity $\boldsymbol{s}_{i_{Tx}}(\omega_{0},\delta(\boldsymbol{x} - \boldsymbol{x}_{i_{Tx}}))$ denotes a source located at Cartesian coordinates $\boldsymbol{x}_{i_{Tx}}$: this variable is zero everywhere except at the (interpolated) grid points of element $i_{Tx}$. For simplicity, we assume that a physical element is described by a single grid point: in this case by $s_{i_{Tx}}(\omega_{0})$ we mean the only non-zero (and generally complex) entry of the matrix  $\boldsymbol{s}_{i_{Tx}}(\omega_{0},\delta(\boldsymbol{x} - \boldsymbol{x}_{i_{Tx}}))$. 
		We indicate by $\boldsymbol{p}_{i_{Tx}}(\omega_{0},\boldsymbol{x}; \boldsymbol{c}_{(k)})$ the numerical solution to the Helmholtz equation at location $\boldsymbol{x}$, at a given frequency  $\omega = \omega_{0}$, and for an underlying speed of sound $\boldsymbol{c}_{(k)}$, when source $i_{Tx}$ is excited ($i_{Tx} = 1,...,N_{Tx}$). This quantity is sometimes called the forward wavefield and it is a complex variable recorded everywhere on the computational grid. We write $p_{i_{Tx},\, j_{Rx}}(\omega_{0}; \boldsymbol{c}_{(k)})$ the restriction of this complex wavefield at the spatial coordinates of receiving element $j_{_{Rx}} = 1,...,N_{Rx}$; this is a complex number.  The forward equation is solved independently for each transmitting element $i_{Tx}$, resulting into $N_{Tx}$ forward wavefields. 

\begin{figure*}[htbp]
	\centering
	 \scalebox{0.5} 
	{\includegraphics{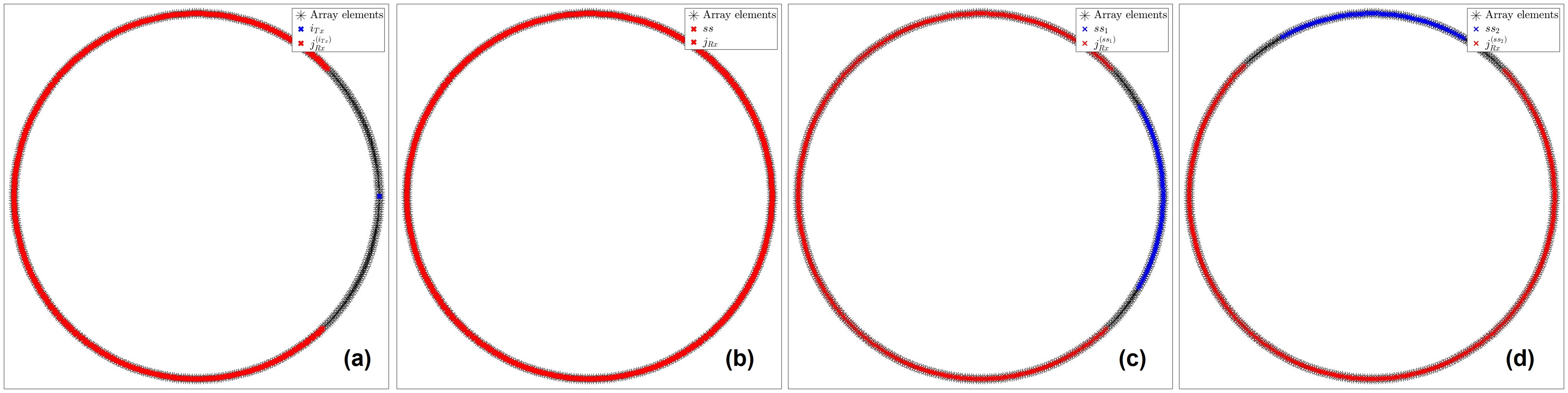}}
	 \caption{\textit{Geometry of a ring-array and selection of transmitters-receivers combinations for deterministic  and stochastic inversions.} Deterministic inversion: single element transmitting, in blue, with highlighted receivers in red (a).  Stochastic inversion: all physical elements  act as transmitters (one super-shot) and as receivers (b),  a subgroup of elements act as transmitters (first super-shot), in blue, with the corresponding receivers  highlighted in red (c),  another subgroup of elements act as transmitters (second super-shot), in blue, with the corresponding receivers  highlighted in red (d).}
	 \label{fig:TxRx} 
	\end{figure*}

The second step is the solution of the so-called adjoint problem. This involves solving another PDE, in this case again the wave equation with the source term given by the residuals, i.e. the difference between the measured data and the simulated data:
\begin{widetext}
	\begin{equation}
	\boldsymbol{A}\left(\omega_{0}, \boldsymbol{x};\,\boldsymbol{c}_{(k)}\right)\, \boldsymbol{q}_{i_{_{Tx}}}(\omega_{0},\boldsymbol{x};\,\boldsymbol{c}_{(k)})\,= -\sum_{j_{Rx}^{(i_{Tx})}}\left[p_{i_{_{Tx}},\,j_{_{Rx}}}(\omega_{0};\boldsymbol{c}_{(k)})-d_{i_{_{Tx}},\,j_{_{Rx}}}^{(obs)}(\omega_{0})\right]^{\star}
	\end{equation}
	\end{widetext}
	where the $\star$ operator is the standard complex conjugate operation on complex numbers. The quantity $\boldsymbol{q}_{i_{_{Tx}}}$ is sometimes called the adjoint wavefieled and it is a complex variable saved everywhere on the computational grid. Having fixed the frequency and the current estimate of the speed of sound, the right hand side of the adjoint equation depends only on the transmitter index $i_{Tx}$. It is implicit that the right hand side is a complex matrix, zero everywhere on the computational grid except at the interpolated locations of the receivers. 
	The expression $j_{Rx}^{(i_{Tx})}$ means that the receiving elements may depend on the transmitting element $i_{Tx}$, in other words the index $j_{Rx}^{(i_{Tx})}$ may run over a (possibly proper) subset among all possible receiving elements $1,...,N_{Rx}$. A notable property of the adjoint equation is that all the elements in $j_{Rx}^{(i_{Tx})}$ fire together (simultaneously). The adjoint equation is solved independently for each transmitting element $i_{Tx}$, resulting into $N_{Tx}$ adjoint wavefields (regardless of how many receiving elements are selected per transmission). At this point it is helpful to revisit the definition of the (single frequency) cost function:
	\begin{widetext}
\begin{equation} \label{Cost}
	C(\omega_{0}; \boldsymbol{c}) = \left\Vert\boldsymbol{d}^{(syn)}(\omega_{0};\boldsymbol{c}) -  \boldsymbol{d}^{(obs)}(\omega_{0})\right\Vert_{2}^{2} = \sum_{i_{Tx}\,=\,1}^{N_{Tx}}\, \sum_{j_{Rx}^{(i_{Tx})}} \left| p_{i_{_{Tx}},\, j_{_{Rx}}}(\omega_{0};\boldsymbol{c}_{(k)})-d_{i_{_{Tx}},\,j_{_{Rx}}}^{(obs)}(\omega_{0})\right|^{2}
\end{equation}
\end{widetext}
where the mapping $i_{Tx}  \rightarrow j_{Rx}^{(i_{Tx})}$ has however to be the same for all iterations, otherwise the cost function is not well defined. In preparation for the results in the next sections, we briefly discuss the case of a ring-array with 512 distinct elements, completely surrounding an object. In analogy with X-ray CT, when transmitting from a single element, one may want to exploit only transmission data in the reconstruction phase. In ultrasound transmission tomography this means to use the information recorded at sensors placed laterally and opposite to a  fixed transmitting element; here we assume an acceptance angle of 270\degree. This means that for $i_{Tx} = 1$, $j_{Rx} = 65, \ldots, 448$, for  $i_{Tx} = 2$, $j_{Rx} = 66, \ldots, 449$, and so on, Fig.  \ref{fig:TxRx} (a).
In particular, for $i_{Tx} = 256$, $j_{Rx} = 1, \ldots, 191$ and $j_{Rx} = 320, \ldots, 512$, in other words one can select receiving elements opposite to $i_{Tx} = 1$ and at the same time exclude some of these from the list of the the receiving elements when $i_{Tx} = 256$ because they do not represent transmission data in the latter case. The ability to surgically select a subgroup of receiving elements according to the location of the transmitting element is an important aspect of the deterministic inversion, which will be evidenced in Section \ref{Results}.

The final step is the calculation of the single frequency, single element gradient of the cost function (\ref{Cost}). This is given by the point-wise multiplication of the adjoint wavefield and the forward wavefield:
\begin{widetext}
	\begin{equation}
	\boldsymbol{\nabla}^{(i_{Tx})}_{\boldsymbol{c}}(\omega_{0},\boldsymbol{x};\boldsymbol{c}_{(k)})\, = \left[\omega_{0}^2\,\,  \boldsymbol{q}_{i_{_{Tx}}}(\omega_{0},\boldsymbol{x};\boldsymbol{c}_{(k)})\,\, \boldsymbol{p}_{i_{_{Tx}}}(\omega_{0},\boldsymbol{x};\boldsymbol{c}_{(k)}\right] * \left(-\,\frac{2}{\boldsymbol{c}_{(k)}^3}\right)
	\end{equation}
	\end{widetext}
	The term in the square brackets represents the gradient with respect to the so-called squared slowness; 	by the derivative chain rule (term in the round brackets), one can calculate the gradient with respect to the speed of sound. The final gradient is obtained by summing over all the transmitting elements and taking the real part 
	\begin{equation} \label{sumGrad}
	\boldsymbol{\nabla}_{\boldsymbol{c}}(\omega_{0},\boldsymbol{x};\boldsymbol{c}_{(k)})\, = \mathfrak{Re}\left[\, \sum_{i_{_{Tx}}\,=\, 1}^{N_{Tx}} \boldsymbol{\nabla}^{(i_{Tx})}_{\boldsymbol{c}}(\omega_{0},\boldsymbol{x};\boldsymbol{c}_{(k)})\,  \right]
	\end{equation}
An update of the speed of the sound is simply obtained through the equation
	\begin{equation}
	\boldsymbol{c}_{(k+1)} = \boldsymbol{c}_{(k)} -  \alpha*\boldsymbol{\nabla}_{\boldsymbol{c}}(\omega_{0},\boldsymbol{x};\boldsymbol{c}_{(k)})
	\end{equation}
		The acoustic impedance matrix $\boldsymbol{A}$ is then re-evaluated, $\boldsymbol{A}\left(\omega_{0}, \boldsymbol{x};\,\boldsymbol{c}_{(k+1)}\right)$, and the previous steps are repeated until convergence. 

The calculation of the gradient requires computing the forward wavefield and the adjoint one as many times as the number of transmitting elements. As wave propagation is linear with respect to the source term, one may want to adopt a strategy where the observed data are summed over all the independent transmissions: this would correspond to the case where all sources would transmit simultaneously. This scenario would be highly beneficial as the forward and the adjoint equations would have to be solved only once, with a dramatic gain in computing times. However, an important aspect of the calculation of the numerical gradient is that the algorithm described above is only valid when one source at the time is active. To illustrate this point, let's consider the case of two distinct sources $i_{1}, i_{2}$ transmitting simultaneously. 
	By linearity, $\boldsymbol{d}^{(obs)}_{i_{2},\,i_{2}} = \boldsymbol{d}^{(obs)}_{i_{1}} + \boldsymbol{d}^{(obs)}_{i_{2}}$, $\boldsymbol{p}_{i_{1},\,i_{2}} = \boldsymbol{p}_{i_{1}} + \boldsymbol{p}_{i_{2}}$ and $\boldsymbol{q}_{i_{1},\,i_{2}} = \boldsymbol{q}_{i_{1}} + \boldsymbol{q}_{i_{2}}$. Omitting unnecessary dependence, and neglecting the term $\left(-2/\boldsymbol{c}_{(k)}^3\right)$, we have 
	\begin{equation}
		\begin{split}
	\boldsymbol{\nabla}^{(i_{1},i_{2})}_{\boldsymbol{c}}  = \omega_{0}^2\, \left[ \boldsymbol{q}_{i_{1},i_{2}}\, \boldsymbol{p}_{i_{1},i_{2}} \right]  \\
	= \omega_{0}^2\, \left[(\boldsymbol{q}_{i_{1}} + \boldsymbol{q}_{i_{2}}) (\boldsymbol{p}_{i_{1}} + \boldsymbol{p}_{i_{2}})  \right]  \\
	= \omega_{0}^2\, \left[\boldsymbol{p}_{i_{1}}\boldsymbol{q}_{i_{1}} + \boldsymbol{p}_{i_{2}}\boldsymbol{q}_{i_{2}} + \boldsymbol{p}_{i_{1}}\boldsymbol{q}_{i_{2}} + \boldsymbol{p}_{i_{2}}\boldsymbol{q}_{i_{1}}\right]  \\
	= \boldsymbol{\nabla}^{(i_{1})}_{\boldsymbol{c}}  + \boldsymbol{\nabla}^{(i_{2})}_{\boldsymbol{c}} + \omega_{0}^2\, \left[\boldsymbol{p}_{i_{1}}\boldsymbol{q}_{i_{2}} + \boldsymbol{p}_{i_{2}}\boldsymbol{q}_{i_{1}}\right]
			\end{split}
	\end{equation}
	The gradient when both sources are active is not the sum of the single source gradients, because of the presence of the cross-terms. This could have been anticipated by looking at the expression for the numerical gradient, as this is not linear with respect to the forward/adjoint wavefields (despite the wave equation and the adjoint equation being both linear with respect to the source term). In other words, computing the gradient with the expression $\boldsymbol{\nabla}^{(i_{1},i_{2})}_{\boldsymbol{c}} = \omega_{0}^2\, \left[ \boldsymbol{q}_{i_{1},i_{2}}\, \boldsymbol{p}_{i_{1},i_{2}} \right] $ wound not return the gradient of the cost function as defined in equation (\ref{Cost}) above, and it would result into solving a different optimization problem.	These considerations motivate the phase-encoding (PE) approach, described in the next section.

	\subsection{Single-frequency Stochastic Optimization}
	
	The stochastic inversion relies on the linearity properties of the wave equation. Our presentation follows \cite{BenHadjAli}. The first step is to form a so-called super-shot, this is a linear superposition of the individual source terms:
	\begin{equation}
	S^{(enc)}(\omega_{0}) = \sum_{i {_{Tx}}\, =\, 1}^{N_{Tx}}\,s^{(enc)}_{i_{Tx}}(\omega_{0}) = \sum_{i {_{Tx}}\, =\, 1}^{N_{Tx}}\, a_{i_{Tx}} \, s_{i_{Tx}}(\omega_{0})
	\end{equation}
	with $s^{(enc)}_{i_{Tx}}(\omega_{0}) =  a_{i_{Tx}}\, s_{i_{Tx}}(\omega_{0})$ and $a_{i_{Tx}}$ a (possibly complex) multiplicative factor, transmitter dependent. This is equivalent to having a single source term $S^{(enc)}$ and corresponds to the case of $N_{Tx}$ (distinct) sources transmitting simultaneously, each weighted (\emph{encoded}) by the term $a_{i_{Tx}}$. By simple linearity arguments, one may define the following quantities
	\begin{equation}
	\boldsymbol{p}^{(enc)}_{i_{Tx}}(\omega_{0},\boldsymbol{x};\boldsymbol{c}_{(k)}) =  a_{i_{Tx}}  \, \boldsymbol{p}_{i_{Tx}}(\omega_{0},\boldsymbol{x};\boldsymbol{c}_{(k)}) 
	\end{equation}
	\begin{equation}
	\boldsymbol{P}^{(enc)}(\omega_{0},\boldsymbol{x};\boldsymbol{c}_{(k)}) =  \sum_{i_{Tx}\, =\, 1}^{N_{Tx}} \boldsymbol{p}^{(enc)}_{i_{Tx}}(\omega_{0},\boldsymbol{x};\boldsymbol{c}_{(k)}) 
	\end{equation}
	that satisfy respectively the two equations:
	\begin{equation}
	\boldsymbol{A}\left(\omega_{0}, \boldsymbol{x};\,\boldsymbol{c}_{(k)}\right)\, \boldsymbol{p}^{(enc)}_{i_{Tx}}(\omega_{0},\boldsymbol{x}; \boldsymbol{c}_{(k)}) =  -s^{(enc)}_{i_{Tx}}(\omega_{0})
	\end{equation}
	\begin{equation}
	\boldsymbol{A}\left(\omega_{0}, \boldsymbol{x};\,\boldsymbol{c}_{(k)}\right)\, \boldsymbol{P}^{(enc)}(\omega_{0},\boldsymbol{x}; \boldsymbol{c}_{(k)}) =  -S^{(enc)}(\omega_{0})
	\end{equation}
	These equations are the analogous of equation (\ref{eq:ForwardD}) in the previous section. By ${P}^{(enc)}_{j_{Rx}}(\omega_{0};\boldsymbol{c}_{(k)}) $ we mean the restriction of the encoded forward wave-field $\boldsymbol{P}^{(enc)}$ at the interpolated grid points for the receiving element $j_{Rx}$: this is a complex number.
The advantage of forming a super-shot is that only one Helmholtz equation has to be solved in order to compare against the recorded data. For this task, one needs to encode the observed data with the same encoding terms:
	\begin{equation}
	\boldsymbol{D}^{(obs,\,enc)}(\omega_{0}) =  \sum_{i_{Tx}\, =\, 1}^{N_{Tx}}  \boldsymbol{d}^{(obs,\, enc)}_{i_{Tx}}(\omega_{0}) = \sum_{i_{Tx}\,=\, 1}^{N_{Tx}} a_{i_{Tx}}  \, \boldsymbol{d}^{(obs)}_{i_{Tx}}(\omega_{0})
	\end{equation}
	The previous quantity is a complex vector of size $N_{Rx} \times 1$; by  $D_{j_{Rx}}^{(obs,\,enc)}(\omega_{0})$ we denote the complex number representative of the encoded measured data at receiver $j_{Rx}$. Accordingly, the single frequency cost function reads as
	\begin{widetext}
		\begin{equation} \label{CostEnc}
	C^{(enc)}(\omega_{0};\boldsymbol{c}) = \left\Vert\boldsymbol{d}^{(syn,enc)}(\omega_{0};\boldsymbol{c}) - \boldsymbol{D}^{(obs,\,enc)}(\omega_{0})\right\Vert_{2}^{2} = \sum_{j_{Rx}\,=\,1}^{N_{Rx}} \left| P_{j_{_{Rx}}}^{(enc)}(\omega_{0};\boldsymbol{c}_{(k)})-D_{j_{_{Rx}}}^{(obs,\,enc)}(\omega_{0})\right|^{2}
	\end{equation}
	\end{widetext}
	and it is now evident that in this case the receivers have to be shared across all the transmitting elements in the super-shot. For a circular geometry, where all the elements transmit and receive, the distinction between transmission data and reflection data is lost, Fig. \ref{fig:TxRx} (b). Thus, with a single super-shot, the residuals and the cost function are well defined only if the individual transmitters in the super-shot share the same receivers.  
	
	The third step is to solve the adjoint problem
	\begin{widetext}
	\begin{equation}
	\boldsymbol{A}\left(\omega_{0}, \boldsymbol{x};\,\boldsymbol{c}_{(k)}\right)\, \boldsymbol{Q}^{(enc)}(\omega_{0},\boldsymbol{x}; \boldsymbol{c}_{(k)}) =-
	\sum_{j_{Rx}\, =\, 1}^{N_{Rx}} \left[P_{j_{Rx}}^{(enc)}(\omega_{0};\,\boldsymbol{c}_{(k)}) - D_{j_{Rx}}^{(obs,\,enc)}(\omega_{0})\right]^{\star}
	\end{equation}
	\end{widetext}
	with the source term being a sum over all shared receiving elements. The fourth step is the calculation of the gradient:
	\begin{widetext}
	\begin{equation}
	\boldsymbol{\nabla_{c}}^{(enc)}(\omega_{0}, \boldsymbol{x}; \boldsymbol{c}_{(k)}) = \mathfrak{Re}\left\{ \left[\omega_{0}^2\, 
	\boldsymbol{Q}^{(enc)}(\omega_{0},\boldsymbol{x}; \boldsymbol{c}_{(k)})\, \boldsymbol{P}^{(enc)}(\omega_{0},\boldsymbol{x}; \boldsymbol{c}_{(k)})\right] * \left(-\,\frac{2}{\boldsymbol{c}_{(k)}^3}\right) \right\}
	\end{equation}
	\end{widetext}
	The main difference with respect to the analogous expression in the previous section is that the sum over the individual transmissions has been replaced by a single point-wise multiplication due to the single super-shot condition.  
	Expanding the point-wise multiplication in the right hand side of the previous expression, we obtain
\begin{multline}
	\boldsymbol{Q}^{(enc)}(\omega_{0},\boldsymbol{x}; \boldsymbol{c}_{(k)})\, \boldsymbol{P}^{(enc)}(\omega_{0},\boldsymbol{x}; \boldsymbol{c}_{(k)}) = \\ 
	\left(\sum_{m\,=\, 1}^{N_{Tx}}q^{(enc)}_{m}(\omega_{0})\right)  \left(\sum_{n\, =\, 1}^{N_{Tx}} p^{(enc)}_{n}(\omega_{0})\right)  \\
	= \left(\sum_{m\, =\, 1}^{N_{Tx}} a^{\star}_{m} q_{m}(\omega_{0})\right) \left(\sum_{n\, =\, 1}^{N_{Tx}} a_{n} p_{n}(\omega_{0})\right) \\
	=  \left(\sum_{m\, =\, 1}^{N_{Tx}} \Vert a_{m} \Vert^{2}\, q_{m}(\omega_{0})\, p_{m}(\omega_{0})\right)  + \\
	 \left(\sum_{\substack{m,\,n =\, 1 \\ (n \neq\, m)}}^{N_{Tx}} a^{\star}_{m} a_{n}\, q_{m}(\omega_{0})\, p_{n}(\omega_{0}) \right)  
\end{multline}

	From the previous expression, it's evident that the encoded gradient is not the sum of the individual (single element) gradients. One has an extra contribution due to the so-called cross terms. Until now, we have not made any assumption about the encoding factors $a_{i_{Tx}}$. We now make the further assumption that these are pure phase terms, $a_{i_{Tx}} = \exp{(i\,\phi_{i_{Tx}}})$. This implies $\Vert a_{i_{Tx}} \Vert^{2} = 1$, so that the first term of the encoded gradient is exactly equivalent to the case of having $N_{Tx}$ transmitting elements firing sequentially (independently) and sharing $N_{Rx}$ receivers. If the phase encoding terms are kept fixed for all iterations, then the cost function (\ref{CostEnc}) still defines a deterministic problem (regardless of the calculation rule for the $a_{i_{Tx}}$), different though from the one defined in equation (\ref{Cost}). If the phase encoding terms are drawn from the same probability distribution at each iteration, then the nature of the cost function becomes \emph{stochastic}, and the optimization problem is in principle different from the deterministic one defined in eq. (\ref{Cost}). The equivalence between the deterministic case and the stochastic one is true if one makes the further assumption that the covariance of the $a_{i_{Tx}}$ is the identity matrix
		\begin{equation} \label{eq:aaa}
	\mathbb{E}[\boldsymbol{a} \boldsymbol{a}^{T}]  = \boldsymbol{I}_{N_{Tx} \times N_{Tx}}
		\end{equation}
	In this case, one can prove \cite{Haber} that  the expectation value of the cross terms in the encoded gradient is zero
				\begin{equation}
	\mathbb{E}_{\boldsymbol{a}} \left[  \left(\sum_{\substack{m,\,n =\, 1 \\ (n \neq\, m)}}^{N_{Tx}} a^{\star}_{m}\, a_{n}\, q_{m}(\omega_{0})\, p_{n}(\omega_{0}) \right) \right] = 0
		\end{equation}
	 This can be achieved by drawing the $\phi_{i_{Tx}}$	from a uniform random distribution (i.e. $\phi_{i_{Tx}}$ is in $[0, 2 \pi ]$ uniformly random, for each transmitting element). For the rest of the paper, we assume that the encoding terms are random phase shifts (PS).
	 
With a randomized inversion, an update of the speed of the sound is simply obtained through the equation
	\begin{equation}
	\boldsymbol{c}_{(k+1)} = \boldsymbol{c}_{(k)} -  \alpha*\boldsymbol{\nabla}^{(enc)}_{\boldsymbol{c}}(\omega_{0},\boldsymbol{x};\boldsymbol{c}_{(k)})
	\end{equation}
	The acoustic impedance matrix $\boldsymbol{A}$ is then re-evaluated, $\boldsymbol{A}\left(\omega_{0}, \boldsymbol{x};\,\boldsymbol{c}_{(k+1)}\right)$, and the previous steps are repeated until convergence. 
	
			\subsubsection{Multiple super-shots} \label{super-shots}
	As mentioned, with a single super-shot, Fig. \ref{fig:TxRx} (b), the difference between reflection data and transmission data is lost. The solution is to build \textit{multiple} super-shots, $ss_{i}$. We denote by $N_{ss}$ the number of super-shots; for simplicity, we assume that all super-shots have the same number of transmitting elements, $N_{Tx}^{(ss_{i})}$, and the same number of receiving elements, $N_{Rx}^{(ss_{i})}$; in Fig. \ref{fig:TxRx} (c) and (d), two distinct super-shots are shown with their corresponding receivers ($N_{Tx}^{(ss_{1})} = N_{Tx}^{(ss_{2})} = 86$ and  $N_{Rx}^{(ss_{1})} = N_{Rx}^{(ss_{2})} = 384$).  The generalization of the previous equations to the multiple super-shots scenario is straightforward. Each super-shot is equivalent to an encoded transmission, i.e. all the physical elements in $ss_{i}$ transmit simultaneously; the different super-shots behave as $N_{ss}$ independent and non-simultaneous transmissions in analogy to the single-element transmissions of the deterministic inversion. 
	In particular, the cost function is the sum over the $N_{ss}$ super-shots cost functions
	\begin{widetext}
		\begin{equation}
	C^{(enc)}(\omega_{0};\boldsymbol{c}) = \sum_{ss_{i} \,=\,1}^{N_{ss}}\,C^{(ss_{i})}(\omega_{0};\boldsymbol{c}) = \sum_{j_{Rx}^{(ss_{i})}\,=\,1}^{N_{Rx}^{(ss_{i})} } \left| P_{j_{Rx}}^{(ss_{i})}(\omega_{0};\boldsymbol{c}_{(k)})-D_{j_{_{Rx}}}^{(obs,\,ss_{i})}(\omega_{0})\right|^{2}
	\end{equation}
	\end{widetext}
	where the index $j_{Rx}^{(ss_{i})}$ now runs over super-shot-dependent receiving elements. The gradient is the sum over all super-shots gradients
	\begin{equation}
	\boldsymbol{\nabla_{c}}^{(enc)} \,=\,\mathfrak{Re}\left[\,  \sum_{ss_{i} \,=\,1}^{N_{ss}}\,\boldsymbol{\nabla_{c}}^{(ss_{i})}\right]
	\end{equation}
	where the super-shots gradients are given by
	\begin{equation}
	\boldsymbol{\nabla_{c}}^{(ss_{i})}\, =\, \left[\omega_{0}^2\, 
	\boldsymbol{Q}^{(ss_{i})}\, \boldsymbol{P}^{(ss_{i})}\right] * \left(-\,\frac{2}{\boldsymbol{c}_{(k)}^3}\right) 
	\end{equation}
	with the super-shot forward and adjoint wavefields satisfying the two equations
		\begin{equation}
	\boldsymbol{A}\, \boldsymbol{P}^{(ss_{i})}\,=\,  -S^{(ss_{i})}(\omega_{0})
	\end{equation}
	\begin{equation}
	\boldsymbol{A}\, \boldsymbol{Q}^{(ss_{i})}\,=\,-
	\sum_{j_{Rx}^{(ss_{i})}\, =\, 1}^{N_{Rx}^{(ss_{i})}} \left[P_{j_{Rx}}^{(ss_{i})}(\omega_{0};\,\boldsymbol{c}_{(k)}) - D_{j_{Rx}}^{(obs,\,ss_{i})}(\omega_{0})\right]^{\star}
	\end{equation}
	and  
		\begin{equation}
S^{(ss_{i})}(\omega_{0})\, = \,  \sum_{i {_{Tx}}\, =\, 1}^{N_{Tx}^{(ss_{i})}}\, a_{i_{Tx}} \, s_{i_{Tx}}(\omega_{0})
	\end{equation}
	This strategy allows to differentiate between reflection data and transmission data for each super-shot, at the cost of solving $N_{ss}$ forward and $N_{ss}$ adjoint equations against one forward and one adjoint equation in the single super-shot case. 
				\subsubsection{Multiple stochastic ensembles} \label{stochastic ensembles}
				The previous discussion has so far assumed that the phase encoding terms are generated once per speed of sound iteration. In order to remove residual cross-talk, it may be beneficial to generate the phase encoding terms \textit{multiple} times per super-shot. We refer to this scenario as multiple stochastic ensembles; we denote by $N_{PE}$ the number of times the phase-encoding vectors $\boldsymbol{a}$ are generated per super-shot. The cost function reads
		\begin{equation}
	C^{(enc)}(\omega_{0};\boldsymbol{c}) \, = \,  \sum_{n_{PE} \,=\,1}^{N_{PE}}\,\left[\sum_{ss_{i} \,=\,1}^{N_{ss}}\,C^{(ss_{i},n_{PE})}(\omega_{0};\boldsymbol{c})\right]
	\end{equation}
where each super-shot cost function is now labeled with an independent realization of the phase-encoding terms, $\boldsymbol{a}^{(n_{PE})}$. The final gradient is simply obtained by summing over all super-shots and over all stochastic ensembles
	\begin{equation}
	\boldsymbol{\nabla_{c}}^{(enc)} \,=\,\mathfrak{Re}\left\{\, \sum_{n_{PE} \,=\,1}^{N_{PE}}\,\left[ \sum_{ss_{i} \,=\,1}^{N_{ss}}\,\boldsymbol{\nabla_{c}}^{(ss_{i}, n_{PE})}\right]\right\}
	\end{equation}
		This technique is effective in removing residual cross-talk because it averages statistically independent stochastic gradients $n_{PE}$ times. The scenario described in this section requires solving $N_{ss} \times N_{PE}$ forward and $N_{ss} \times N_{PE}$ adjoint equations. The impact of multiple super-shots in combination with the notion of stochastic ensembles will be highlighted in Section \ref{Results}.
				
		\subsection{Multi-frequency Optimization}
	The generalization to a multi-frequency strategy is straightforward. At the beginning of each iteration, the  acoustic impedance matrices are assembled at the $N_{f}$ discrete frequencies: $\boldsymbol{A}\left(\omega_{1}, \boldsymbol{x};\,\boldsymbol{c}_{(k)}\right), \ldots, \boldsymbol{A}\left(\omega_{N_{f}}, \boldsymbol{x};\,\boldsymbol{c}_{(k)}\right)$. This step is the same for both inversion algorithms. 	The additional reconstruction steps are run independently at each discrete frequency (in parallel or sequentially, depending on the computing resources), the single-frequency gradients are summed
			\begin{equation}
\boldsymbol{\nabla_{c}}(\omega_{1},\ldots,\omega_{N_{f}}, \boldsymbol{x}; \boldsymbol{c}_{(k)}) = \sum_{\omega_{i}\, =\, 1}^{N_{f}}  \boldsymbol{\nabla_{c}}(\omega_{i}, \boldsymbol{x}; \boldsymbol{c}_{(k)})
	\end{equation}
		\begin{equation}
\boldsymbol{\nabla_{c}}^{(enc)}(\omega_{1},\ldots,\omega_{N_{f}}, \boldsymbol{x}; \boldsymbol{c}_{(k)}) = \sum_{\omega_{i}\, =\, 1}^{N_{f}}  \boldsymbol{\nabla_{c}}^{(enc)}(\omega_{i}, \boldsymbol{x}; \boldsymbol{c}_{(k)})
	\end{equation}
and the speed of sound is updated
\begin{equation}
	\boldsymbol{c}_{(k+1)} = \boldsymbol{c}_{(k)} -  \alpha*\boldsymbol{\nabla}_{\boldsymbol{c}}(\omega_{1},\ldots,\omega_{N_{f}},\boldsymbol{x};\boldsymbol{c}_{(k)})
	\end{equation}
\begin{equation}
	\boldsymbol{c}_{(k+1)} = \boldsymbol{c}_{(k)} -  \alpha*\boldsymbol{\nabla}^{(enc)}_{\boldsymbol{c}}(\omega_{1},\ldots,\omega_{N_{f}},\boldsymbol{x};\boldsymbol{c}_{(k)})
	\end{equation}
The  acoustic impedance matrices are then re-assembled at the $N_{f}$ discrete frequencies, $\boldsymbol{A}\left(\omega_{1}, \boldsymbol{x};\,\boldsymbol{c}_{(k+1)}\right), \ldots, \boldsymbol{A}\left(\omega_{N_{f}}, \boldsymbol{x};\,\boldsymbol{c}_{(k+1)}\right)$, and all the steps are repeated until convergence. 

Finally, one may wish to invert sequentially over multiple frequency bandwidths, each containing one or more discrete frequencies. In this case, the cost function, deterministic or stochastic, is re-defined for each bandwidth, in other words each bandwidth defines an independent optimization problem.

	\section{Results} \label{Results}
	To validate our reconstruction algorithm on synthetic data, we have run 2D and 3D acoustic modeling following the nominal design parameters of the system described in \cite{DuricSD} and in \cite{DuricPMB}. The latter is a circular array with 2048 physical elements; of these only 1048 are used to transmit and receive. The array has a radius of 110 mm and transmits at a central frequency of 2.75 MHz. The breast is scanned with a 3 mm sampling in elevation, with a tentative slice thickness (beam-width) of 3 mm at the central frequency. The focal depth of the focused transmission in elevation has not been publicly disclosed. Data are recorded for 176 $\mu s$, sampled at 12 MHz with a 14 bits ADC and then interpolated to 16 bits; only frequencies larger than 300-400 kHz are above noise floor. The system has 512 channels. Recently, two experimental datasets collected with the described device have been made publicly available in \cite{DuricLast} (512 x 512 combinations of transceivers have been shared); we refer to these as Malignancy and Cyst. For the numerical studies, we limit ourselves to the case of a ring-array with 512 physically distinct elements and a radius of 110 mm. In the following sections, we consider both 2D forward data, in the frequency-domain (FD) or in the time-domain (TD), and 3D forward data (TD). In the 2D case, array elements are modeled as perfectly spherical emitters, i.e. the Cartesian coordinates of a single element are mapped to one grid-point. In the 3D case, array elements are modeled by a single arc-shaped segment with a focal depth of 55 mm and beam-width of 3 mm (for alternative transducer modeling see \cite{Anastasio3}). In the case of a FD2D model, forward data have been modeled with a 9 points finite difference scheme at a few discrete frequencies; TD forward data have been modeled with the popular k-Wave toolbox. Also, since frequencies larger than 1 MHz are ignored for the purpose of image reconstruction, we have considered the case of a wide-band pulse at the central frequency of 1 MHz. The parameters for the forward simulations are summarized in Table \ref{Tab:Tab1}.

\begin{figure*}[htbp]
	\centering
	 \scalebox{0.5} 
	{\includegraphics{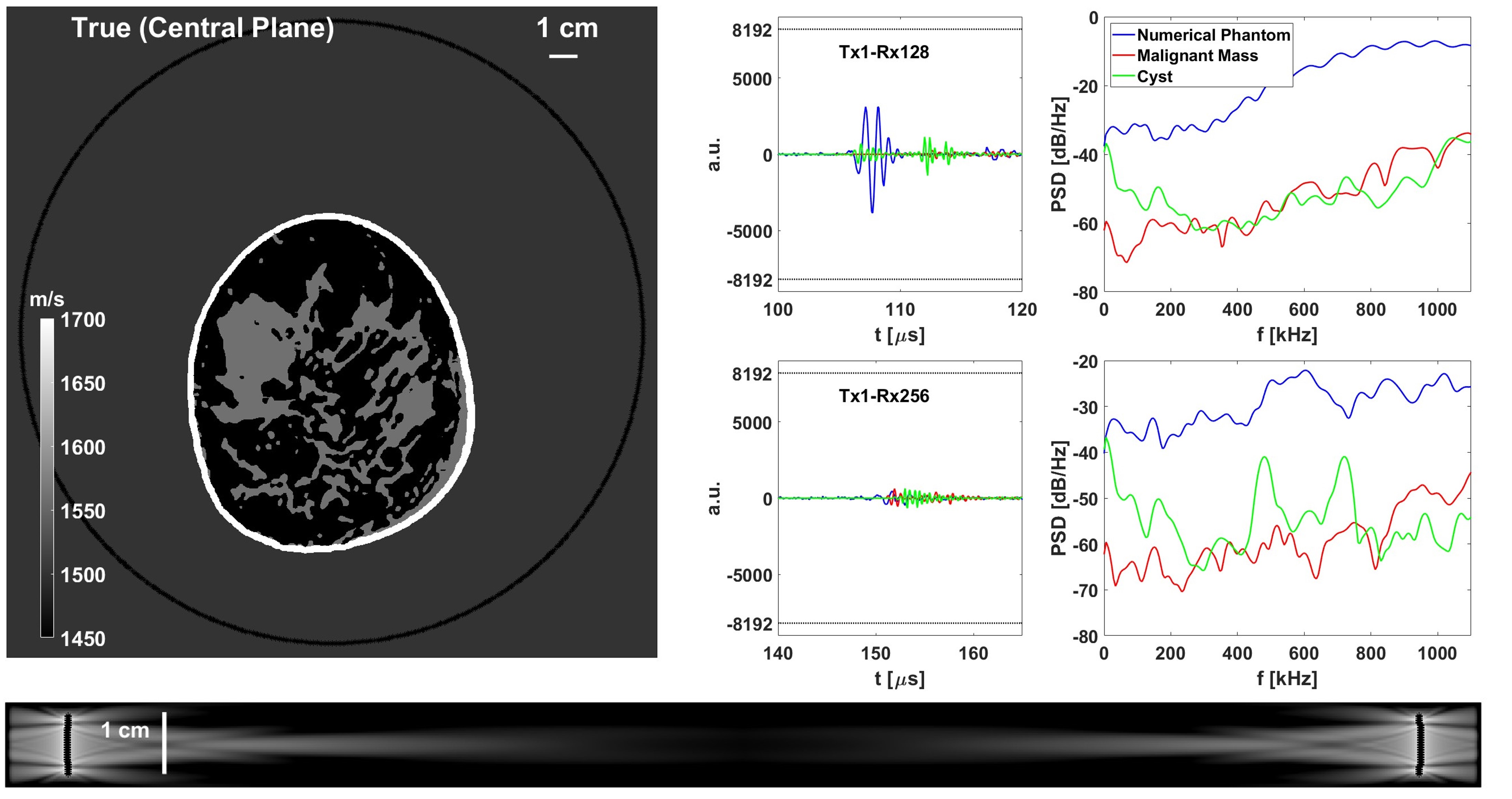}}
	 \caption{\textit{Numerical Phantom, Transducers Geometry and comparison between syntehtic data and experimental raw data power spectral densities.}}
	 \label{fig:PhantomGeometry} 
	\end{figure*}

The numerical phantom employed for the numerical studies is described in \cite{Anastasio4}. This is representative of an anatomical breast with fat tissue (speed of sound 1450 m/s), dense tissue (speed of sound 1540 m/s) and skin layer (speed of sound 1700 m/s). The breast is immersed in a water bath with a speed of sound of 1500 m/s. The phantom, the geometry of the ring-array and a comparison between (TD3D) synthetic data and experimental data are shown in Fig. \ref{fig:PhantomGeometry}. In particular, TD3D forward data have been scaled to have similar amplitudes as transmission data of the experimental datasets; data have then been capped to the saturation limit of a 14 bits ADC, converted to 16 bits and re-cast to single precision for the purpose of image reconstruction. Gaussian noise has been added with a power that only the frequencies larger than 300-400 kHz are above noise floor; in combination with the scaling of the data, this preserves the system SNR of the device (below 1 MHz). By system SNR, we mean the height of the signal level at the relevant frequencies over the noise floor in the frequency domain, in other words the power spectral density (PSD). However, tissue attenuation has not been included in the forward modeling. 

In the following sections, we present the results of the two inversion algorithms described above, by progressively considering scenarios that are more and more representative of the experimental system. The parameters for the image reconstruction phase are summarized in Table \ref{Tab:Tab2}. The abbreviations for the reconstructed images and the cost functions are summarized in Table \ref{Tab:Tab3}; the individual labels are introduced and explained in the relevant sections below. 
		\begin{table*}
		\centering
				\caption{Parameters for forward simulations.}

		\begin{tabular}{|p{90pt}|p{120pt}|p{120pt}|p{120pt}|}
		\hline
		  & FD2D & TD2D & TD3D \\
		\hline
		\hline
		Grid Size & 360 x 360 & 1200 x 1200  & 1200 x 1200 x 68 \\
             Pixel Size & 0.8 mm  & 0.2 mm  & 0.2 mm \\
		Time Step &  - & 83.3 ns (12 MHz) & 83.3 ns (12 MHz) \\
		Time Samples & - & 2400 & 2400  \\
		Frequency & 100-200-300 kHz & 1 MHz (central, pulsed) & 1 MHz (central, pulsed) \\
		Focal Depth & - & -  & 55 mm \\
	      Beam-width & - & - & 3 mm \\
		Ring Radius & 110 mm & 110 mm  & 110 mm \\
		Array Elements & 512 & 512 & 512 \\
		\hline
		\end{tabular}
		\label{Tab:Tab1}
		\end{table*}	

				\begin{table*}
					\centering 
		\caption{Parameters for frequency-domain reconstruction.}
		\label{Tab:Tab2}
		 \begin{tabular}{|p{120pt}|p{60pt}|p{50pt}|p{140pt}|}
		\hline
		 Inverted Frequencies & Grid Size & Pixel Size & Forward Model\\
		\hline
		\hline
		100-200-300 kHz & 360 x 360 & 0.8 mm & FD2D\\
		100-200 kHz (BW1) & 280 x 280 &  1 mm & TD2D, TD3D, Experimental Data\\
		300 kHz (BW2) & 350 x 350 &  0.8 mm & TD2D, TD3D, Experimental Data\\ 
	      400 kHz (BW3) & 466 x 466 &  0.6 mm  & TD2D, TD3D, Experimental Data\\ 
		500-600-700 kHz  (BW4) & 700 x 700 &  0.4 mm & TD2D, TD3D, Experimental Data\\ 
		800-900-1000 kHz  (BW5) & 875 x 875 &  0.32 mm & TD2D, TD3D, Experimental Data\\ 
		\hline
		\multicolumn{4}{p{420pt}}{All (stochastic) deterministic inversions have been performed with a (stochastic) gradient descent algorithm with inexact line search.}\\
		\end{tabular}
		\end{table*}
						\begin{table*}
					\centering 
		\caption{List of abbreviations.}
		\label{Tab:Tab3}
		\setlength{\tabcolsep}{3pt}
		 \begin{tabular}{|p{120pt}|p{250pt}|}
		\hline
		 Abbreviation & Definition \\
		\hline
		\hline
		BW & reconstruction bandwidth \\
		PE-R & stochastic reconstruction with phase-encoding (Radamacher) \\
	   PE-PS & stochastic reconstruction with phase-encoding (phase-shifts) \\
   	   NSS & number of super-shots \\
		WIN & window along receivers \\ 
	   NOISE & noisy data (simulations only) \\ 
		SMOOTH & smoothing filter applied on the gradient \\ 
		\hline
		\multicolumn{2}{p{400pt}}{All reconstructed images and cost functions whose labels (do not) include the PE- prefix are the result of a (deterministic) stochastic inversion.}\\
		\end{tabular}
		\end{table*}
	
	\subsection{Inverse Crime (FD2D-FD2D)} \label{InvCrime}
	In this section, we test our implementation of the inversion algorithms for the inverse crime case. Specifically, the forward problem has been solved on a grid of size 360 x 360 pixels, with a pixel size of 0.8 mm (this phantom has been obtained by interpolating the phantom described above). The Helmholtz equation has been solved at three discrete frequencies 100 - 200 - 300 kHz. The corresponding source terms have been extracted from the FFT of the same excitation pulse used for the TD simulations; the same source waveform has been employed for all transmitters. A full data set is a complex variable of size 3 x 512 x 512, where the first dimension represents the frequencies, the second dimension represents the receivers and the third dimension represents the transmitters. The reconstruction grid is the same as the forward one; we also assume perfect knowledge of the transmission pulse. A homogeneous velocity model has been employed as initial guess (1500 m/s). Gradient descent with inexact line search tracking the behavior of the cost function has been employed (max 5 line searches). The step size is initially set according to the rule:
	$$
	\alpha = \frac{40}{\max{[-\boldsymbol{\nabla}_{\boldsymbol{c}}(\boldsymbol{c}_{(1)})]}} 
	$$ 
	where the expression $\max{[-\boldsymbol{\nabla}_{\boldsymbol{c}}(\boldsymbol{c}_{(1)})]}$ refers to the maximum value of the descent direction (the latter being equal to the sum over all the single frequency gradients, up to a sign), at iteration 1. The scaling factor 40 follows form the observation that such a rule updates an initial homogeneous velocity model (1500 m/s) into a velocity map with a maximum value equal to 1540 m/s, average value for the the speed of sound in soft tissue. The step size is then tuned during the line search iterations and it is reset to its original value at the beginning of each iteration. The deterministic inversion has been run by inverting for the three discrete frequencies simultaneously, for a total of 200 iterations.  The randomized inversion has been run by inverting for the three discrete frequencies simultaneously, for a total of 1000 iterations. For the randomized inversion, a similar strategy has been employed with the extra caveat that the phase encoding terms change at each iteration \emph{but} they are fixed to their (speed of sound) iteration value within the line search loop. In both cases, the speed of sound is updated on the entire grid, in other words the reconstruction FOV (field of view) coincides with the computational grid.  All combinations of transmitters-receivers have been employed for both inversions. 
		\begin{figure*}[htbp]
	\centering
	\scalebox{0.5} 
	{\includegraphics{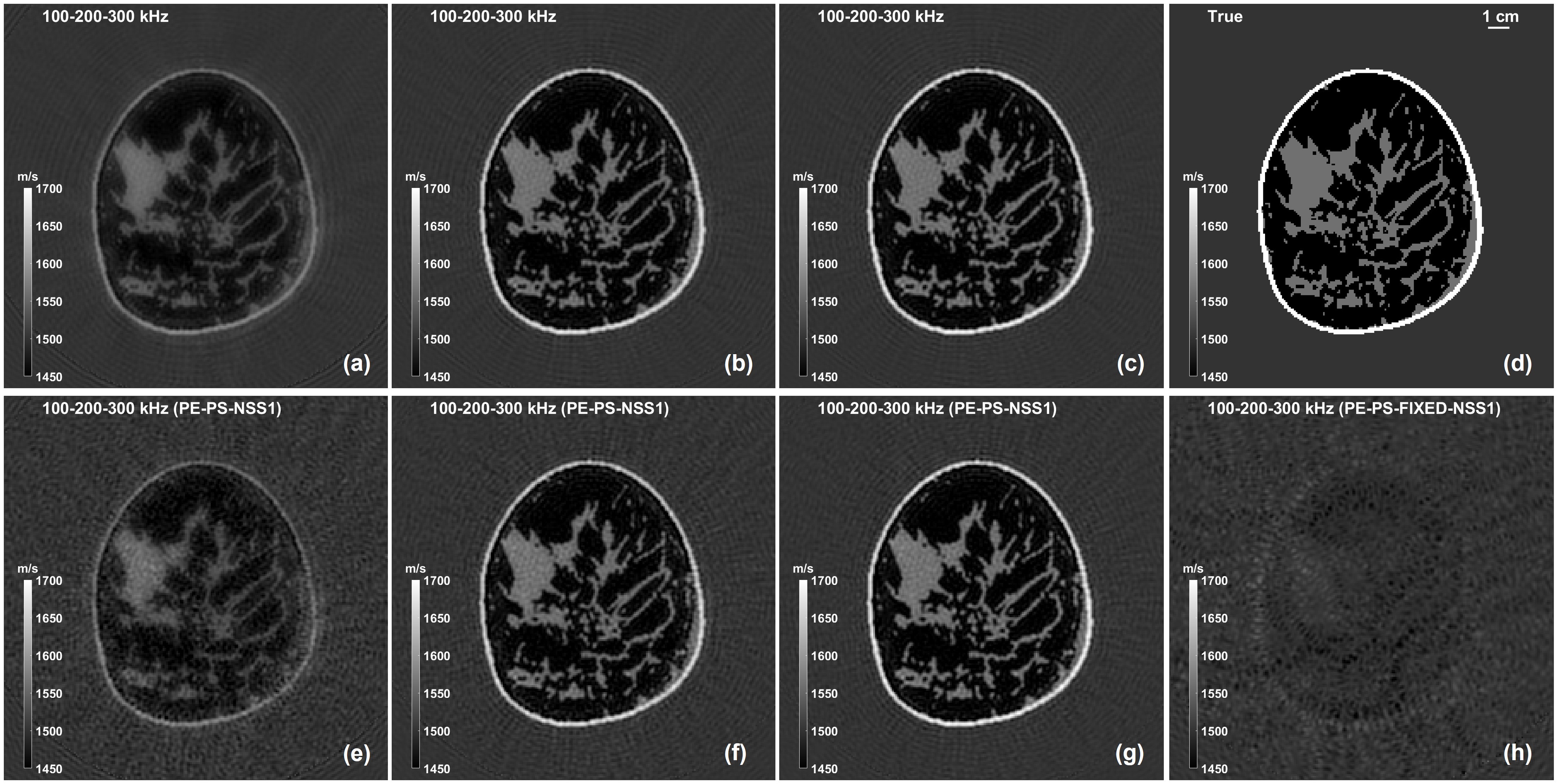}} 
	\centering
	\scalebox{0.5} 
	{\includegraphics{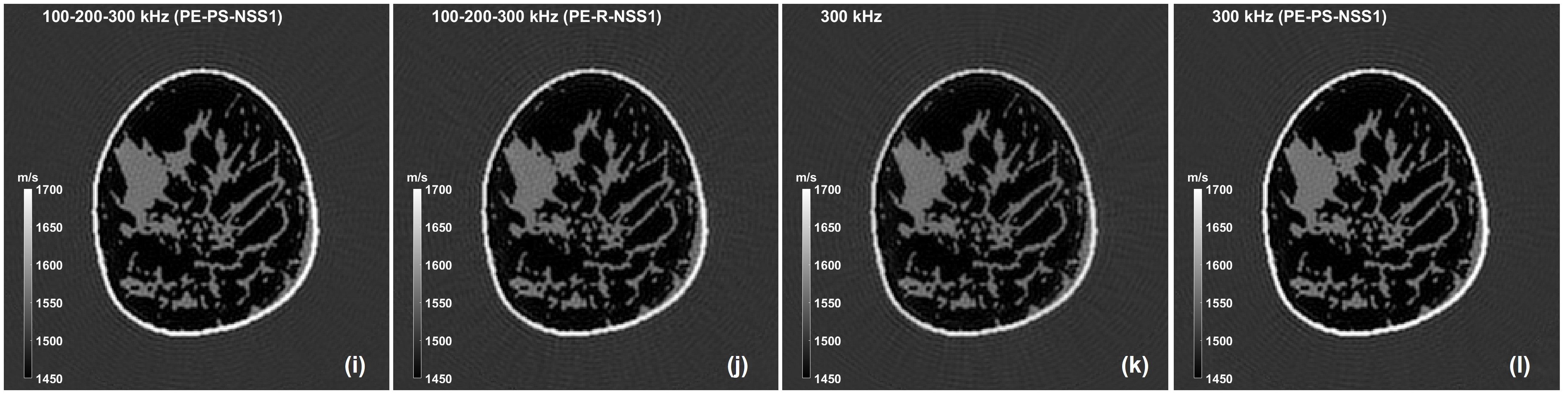}}
	\centering
	\scalebox{0.5} 
	{\includegraphics{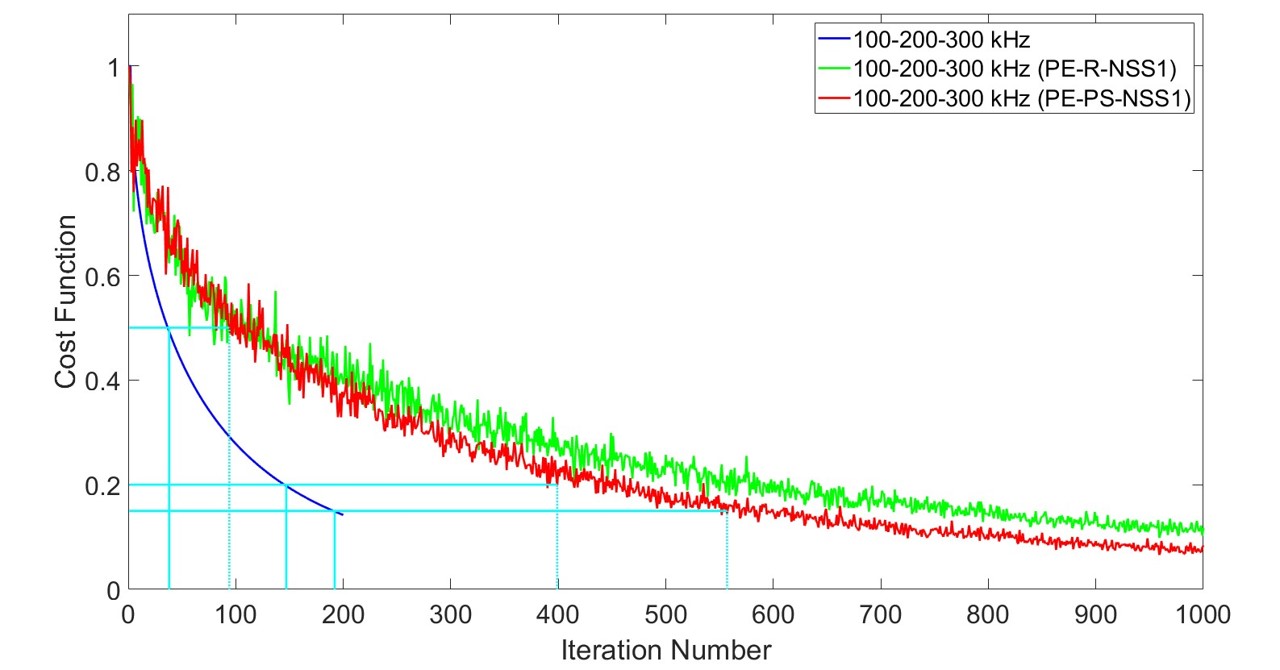}} 
	\caption{\textit{Inverse Crime (Forward FD2D - Reconstruction FD2D). Deterministic and stochastic inversion. Reconstructed images and cost functions.} First row: deterministic inversion in correspondence of cost function values of 0.5 (a), 0.2 (b) and 0.1 (c) respectively, true object (d). Second row: stochastic inversion in correspondence of cost function values of 0.5 (e), 0.2 (f) and 0.1 (g) respectively, stochastic inversion with fixed phase-encoding vectors after 1000 iterations (h). Third row: stochastic inversions with phase-shifts (i) and Radamacher (j) encoding respectively after 1000 iterations, deterministic (k) and stochastic (l) inversion at a single frequency of 300 kHz (after 200 and 1000 iterations respectively). }
	\label{fig:InverseCrime1}
	\end{figure*}
	The results of these numerical studies are summarized in Fig. \ref{fig:InverseCrime1}. In particular, intermediate results corresponding to (normalized) cost function values of 0.5, 0.2 and 0.1 are shown, for the deterministic inversion in Fig. \ref{fig:InverseCrime1} (a) (b) (c), and for the stochastic inversion in Fig. \ref{fig:InverseCrime1} (e) (f) (g), respectively. As expected, image quality improves iteration after iteration, and, as known, it takes many more iterations for the randomized inversion (PE-PS) to reach comparable image quality. In all cases, both inversions show a wavy pattern on top of the image, due to the low-frequency regime. The figure also shows the result of a randomized inversion by keeping fixed the phase encoding terms for all iterations (PE-PS-FIXED), Fig. \ref{fig:InverseCrime1} (h): in this case, the inversion fails, and even after hundreds of iterations the algorithm it not capable of updating the speed of sound. 
	 This proves numerically that the theoretical equivalence between the deterministic inversion and the randomized one has to be interpreted in a probabilistic sense as described above, i.e. by evaluating the cost function at each iteration step with a new realization of the encoding vector. It is only in this sense, and after many iterations and many realizations of the encoding vectors, that the cross-talk among all the combination of transmitters-receivers can be removed. The variability of the cost function, i.e its apparent increase in the stochastic case, is indeed due to the fact that the phase encoding terms are randomly generated at each iteration, hence the cost function is re-evaluated with a different scaling of the data. To test the impact of the probability distribution, we have considered two cases: (PS), random phase-shifts as defined above, Fig. \ref{fig:InverseCrime1} (i), and (R), for Radamacher distribution equivalent to the so-called polarity case in randomized time-domain inversions \cite{Krebs} (the latter being obviously a very special case of phase shifts), Fig. \ref{fig:InverseCrime1} (j). The cost functions for both encoding schemes are similar; however, the cost function for the (PS) case seems to bound the cost function for the (R) case from below. This may or may not be phantom-dependent; for the remaining of this paper, we only consider the (PS) scheme, which is more general. Finally, we test the reconstruction at a single frequency of 300 kHz, deterministic Fig. \ref{fig:InverseCrime1} (k) and  stochastic Fig. \ref{fig:InverseCrime1} (l). This test is motivated by the fact that current devices on the market are not capable, by design, of recording frequencies lower than 300-400 kHz. In both cases, image quality is comparable to the respective three frequencies inversion. As solving the Helmholtz equation is frequency dependent and computationally intensive, one may then be tempted to adopt a strategy where inversion starts at higher frequencies. This strategy is known to fail because of the non-convexity of the cost function defined above: in FWI \cite {VirieuxOperto} this phenomenon is known as cycle skipping or phase wrapping \footnote{There are instances of images corrupted by the cycle skipping artifacts even when using a travel time tomography initial velocity model, see Fig. 2(d) in \cite{DuricCS}. It is unclear to the author how to protect waveform inversion from such events in a \textit{robust} way. Recently, a promising approach has been described in \cite{Wu}; these results, however, are limited to numerical studies in 2D, with full inverse crime, so further investigations are needed.}. For a breast slice of this size, with the assumed speed of sound distribution and with the given geometry, this strategy is successful; in particular the randomized inversion seems to preserve the robustness to cycle skipping artifacts when the deterministic one is immune to it. For a counter-example, and for evidence that the randomized inversion exhibits the same cycle skipping artifacts as the deterministic inversion when the latter exhibits these, the reader is referred to the brief discussion in Appendix \ref{AppendixA}.
		
		\subsection{Without Inverse Crime (TD2D-FD2D)}
	In this section, we present the results for the case where no inverse crime is committed. These results are derived assuming the same physical model (i.e. acoustics in 2D) for the forward and for the reconstruction phases, but the numerical models are different (TD vs FD), as well as the size of the grids and the value of the pixels. Forward data have been generated in time-domain in 2D (TD2D in Table \ref{Tab:Tab1}). The time-domain data set is a real variable of size 2400 x 512 x 512, where the fist dimension represents the time samples (12 MHz sampling frequency), the second dimension represents the receivers and the third dimension represents the transmitters. Data in the frequency domain are obtained after taking an FFT; 10 discrete frequencies are extracted at the corresponding frequency bins, 100 - 1000 kHz, with a step of 100 kHz. The resulting data set is a complex variable of size 10 x 512 x 512. The inversion phase exploits a multi-scale, multi-frequency strategy, where groups of frequencies are sequentially inverted on grids of the same size but progressively smaller pixel size, Table \ref{Tab:Tab2}. Frequencies are grouped in 5 bandwidths, each arbitrarily containing one, two or three frequencies. To minimize inverse crime, the smallest pixel size, 0.32 mm, has been chosen larger than the forward pixel size, 0.2 mm, but as small as possible compatibly with a maximum inverted frequency of 1 MHz. The Cartesian coordinates of the transceivers are assumed to be known and the initial velocity model is again a homogeneous map (1500 m/s); we don't consider these assumptions falling under the inverse crime case as these can be estimated precisely in an experimental setting too (discussed below). The estimation of the source waveform is as follows. The source term is initially described by the same complex number $s = 1 + 1\,i$ for all the elements. For the deterministic inversion, simulated data obtained with this source term are then scaled to match amplitude and phase
of the observed (forward) data, \emph{for each transmitting element} (equation (17) in \cite{Pratt}, see also discussion in \cite{VirieuxOperto}). This process is repeated at the beginning of each iteration; rigorously speaking, this introduces another (complex) parameter in the cost function, parameter that it is updated at each iteration. For the randomized inversion, as in the inverse crime discussion, we only consider the case of a single super-shot ($N_{ss} = 1$). In this case, the same complex number $s = 1 +1\,i$ is employed for all transmitters and the simulated data are then scaled to match amplitude and phase of the encoded forward
data at \emph{all} the 512 elements; this process is repeated at the beginning of each iteration. The minimization strategy per bandwidth is the same as in the previous section. For each bandwidth BW, the initial step size is set according to the rule
	$$
	\alpha^{(BW)} = \frac{40}{\max{[-\boldsymbol{\nabla}_{\boldsymbol{c}}^{(BW)}(\boldsymbol{c}_{(1)})]}} 
	$$ 
	where $\boldsymbol{\nabla}_{\boldsymbol{c}}^{(BW)}(\boldsymbol{c}_{(1)})$ is the gradient with respect to the speed of sound, at iteration 1 in bandwidth BW. Line searches are run in each bandwidth. When inverting on a higher frequency bandwidth (finer pixel size), the speed of sound is first interpolated onto the new grid.  All combinations of transmitters-receivers have been employed for both inversions.
	
\begin{figure*}[htbp]
	\centering 
	 \scalebox{0.5} 
	 {\includegraphics{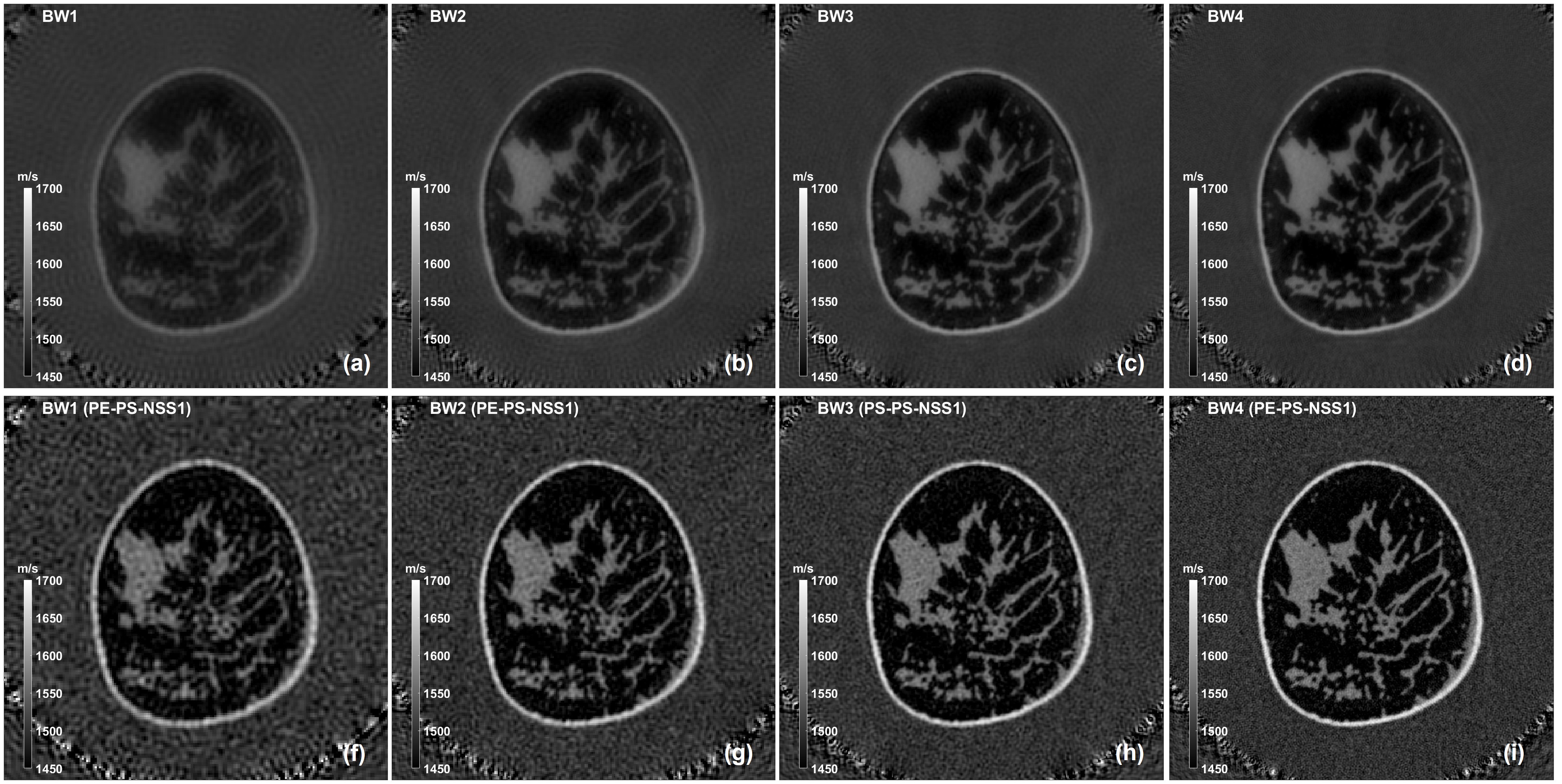}} 
	 \centering 
	 \scalebox{0.502} 
	 {\includegraphics{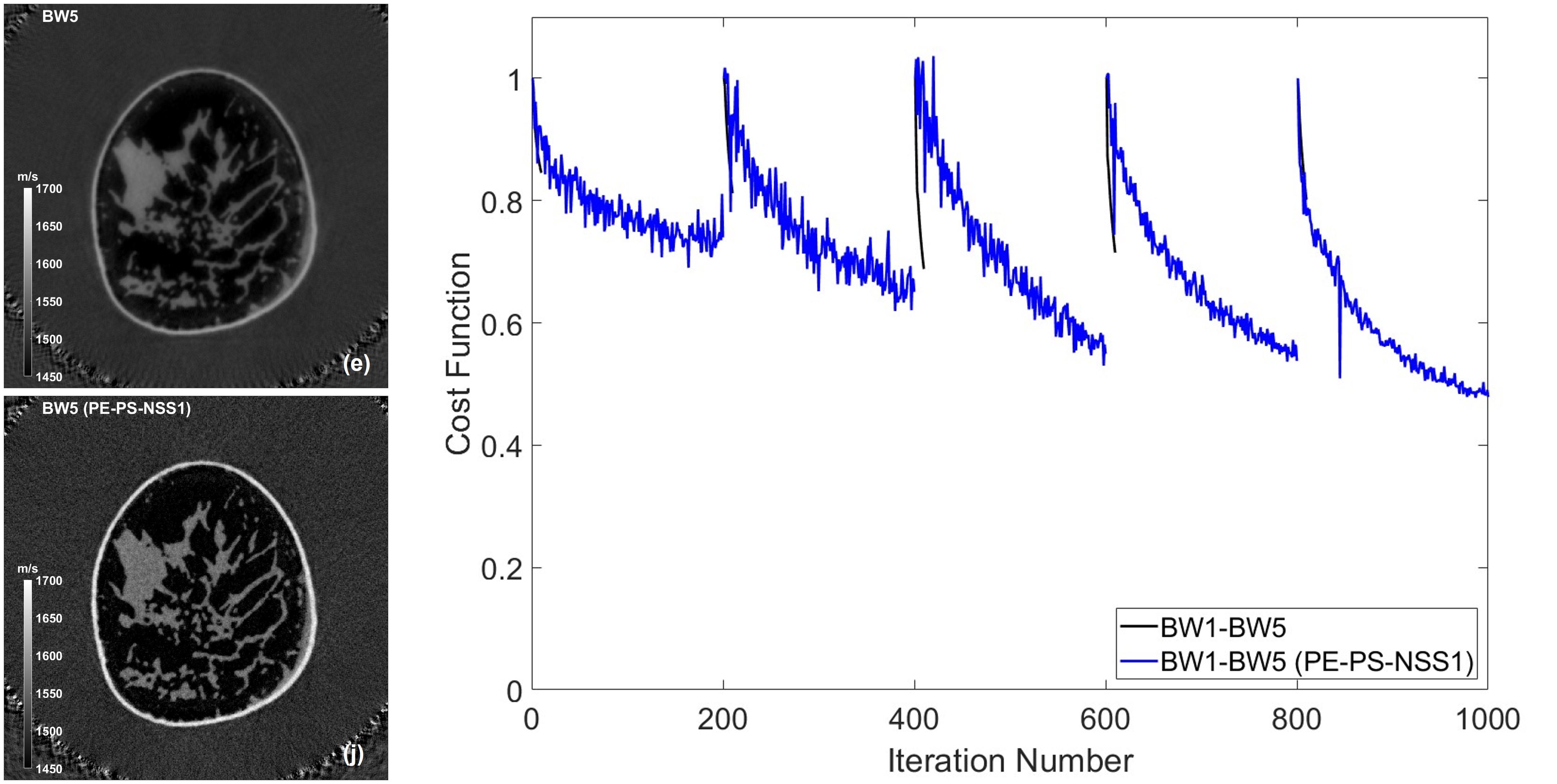}} 
	 \caption{\textit{Without Inverse Crime (Forward TD2D - Reconstruction FD2D). Deterministic and stochastic inversion. Reconstructed images and cost functions.} Multi-scale, multi-frequency image reconstruction: deterministic inversions (a), (b), (c), (d), (e), stochastic inversions (f), (g), (h), (i), (j). Deterministic (stochastic) inversions are run for 10 (200) iterations per bandwidth.}
	 \label{fig:NoInverseCrime2D} 
	\end{figure*}

	Deterministic inversions are run for 10 iterations per bandwidth, randomized ones are run for 200 iterations for bandwidth. Results are shown in Fig. \ref{fig:NoInverseCrime2D} for each bandwidth, deterministic (a)-(e), stochastic (f)-(j). All cost functions are normalized to their respective initial value in each frequency bandwidth.	As expected, image quality improves inverting towards higher frequencies; the previously observed wavy pattern on top of the images disappears in BW3 (400 kHz), and this is true both for the deterministic inversion and the randomized one. The final randomized image in BW5 shows  speed of sound values closer to the true ones with respect to the  corresponding deterministic image: this is due to that fact that the deterministic inversion is under-converged, as evidenced by the graph of the cost function in BW5. This phenomenon proves once more the inherent nature of iterative methods, i.e. that image quality depends on the number of iterations. 
	On the other hand, the stochastic inversion in BW5, Fig. \ref{fig:NoInverseCrime2D} (j), shows a \emph{texture} on top of the reconstructed (average) values of the speed of sound of the different tissues, texture that is \emph{not} present in the true phantom, Fig. \ref{fig:InverseCrime1} (d), and that is \emph{not} present in the deterministic inversion, Fig. \ref{fig:NoInverseCrime2D} (e). This random pattern is due to the stochastic nature of the \emph{phase encoding signal} (although random, we name the latter signal to differentiate it from the random noise due to the analog and digital stages in the receiver electronics, to be considered below). Although in a medical image the pixel values of a given tissue always show a certain variability, an additional and evident texture on top of the true pixel values may be read as a sign of an underlying pathology and eventually lead to a wrong diagnosis. This random pattern is evident even after running hundreds of iterations per bandwidth; and it doesn't appear to decrease when inverting towards higher frequencies. Hence, a way to mitigate the phase encoding signal is needed in order to restore an image quality comparable to the one of a deterministic inversion. We address this in the next section.
	
	\subsection{Model Mismatch (TD3D-FD2D)}
	We now consider the more realistic scenario where the forward model is 3D (TD3D in Table \ref{Tab:Tab1}). The results in this section are derived assuming different physical (and transducers) models for the forward and for the reconstruction phases, acoustics in 3D in time-domain and acoustics in 2D in the frequency domain respectively. The main difference is not insomuch in the TD forward model vs the FD reconstruction model, but it lies in the well known fact that the physics of wave propagation is different in 3D and in 2D (and, more generally, even spatial dimensions vs odd spatial dimensions). In fact, the Green's functions in 2D and in 3D have different physical properties: a long-tail exists in the time-domain signal in the 2D case compared to the 3D ones, on top of a different scaling of the pressure in 2D vs 3D. Together with the design choice of having a 2D system which introduces out-of-plane scattering that cannot be resolved by a planar system, this will impact the quality of any 2D reconstruction, deterministic or stochastic. We show here that such an impact is the same for both methods.

As in the previous section, the time-domain data set is a real variable of size 2400 x 512 x 512. Data in the frequency domain are obtained after taking an FFT; the resulting data set is a complex variable of size 10 x 512 x 512. We discuss the two inversions separately. 
		\subsubsection{Deterministic Inversion}
\begin{figure*}[htbp]
	\centering 
	 \scalebox{0.5} 
	 {\includegraphics{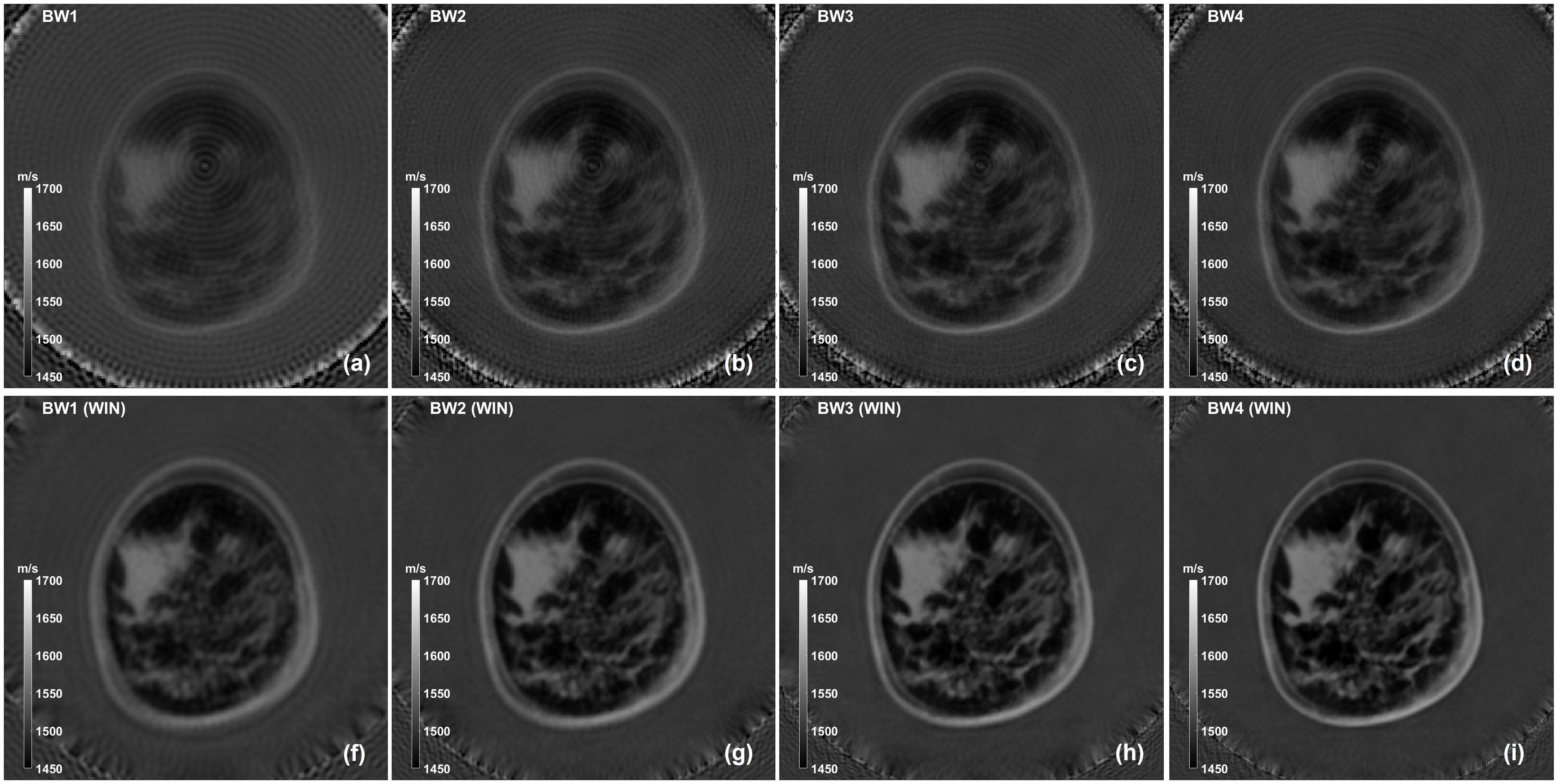}} 
	 \centering 
	 \scalebox{0.5} 
	 {\includegraphics{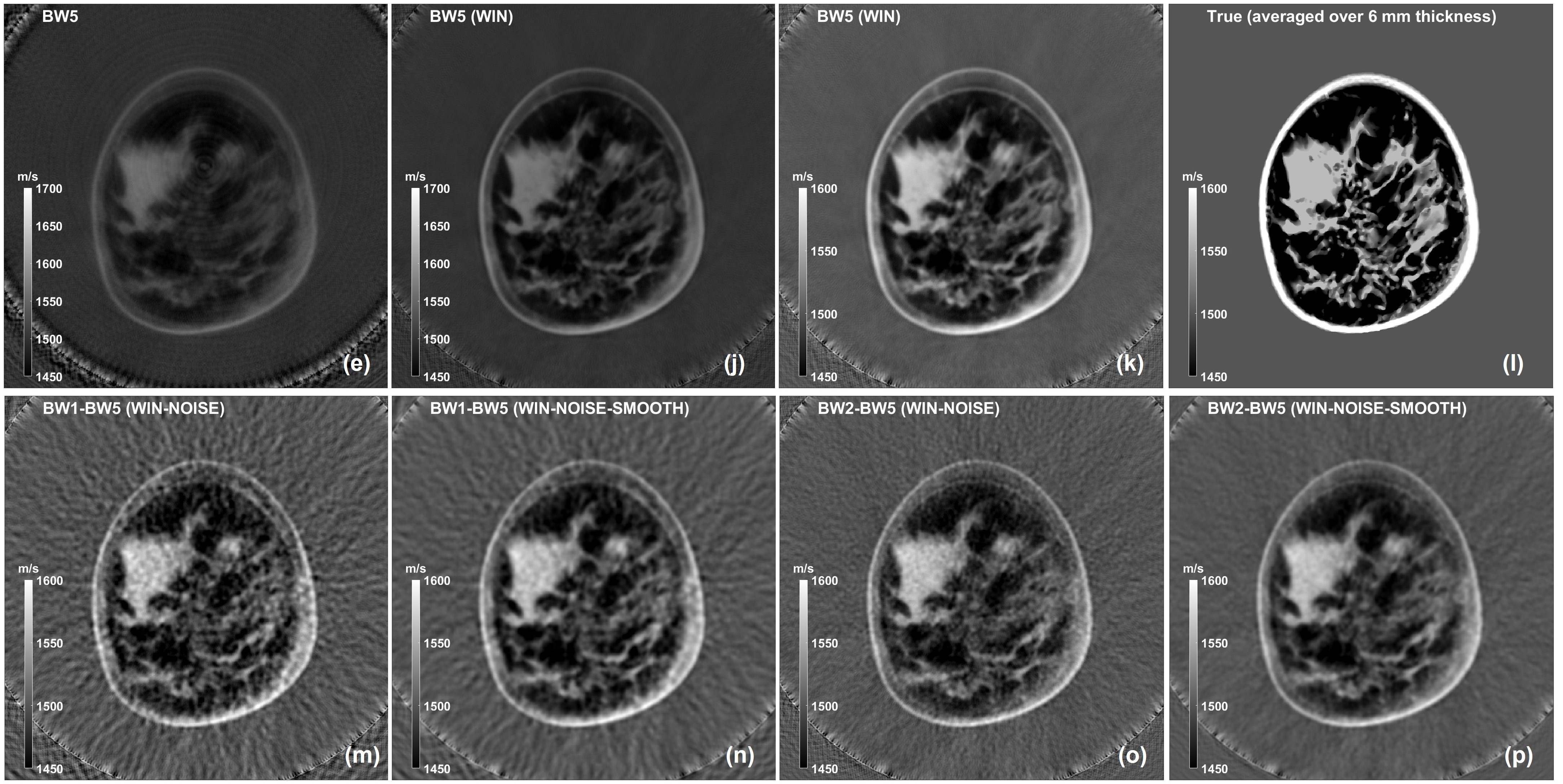}} 
	 \caption{\textit{Model mismatch  (Forward TD3D - Reconstruction FD2D). Deterministic inversion. Reconstructed images  (noiseless and noisy data).} Multi-scale, multi-frequency image reconstruction (noiseless data): deterministic inversions with all combinations of transceivers (a), (b), (c), (d), (e), deterministic inversions with windowed data (f), (g), (h), (i), (j), final image displayed on a different range (k). Central slice of the true phantom averaged over a thickness of 6 mm (l). Multi-scale, multi-frequency image reconstruction (noisy data): reconstruction from BW1 to BW5 with windowing of the data (m), reconstruction from BW1 to BW5 with windowing of the data and smoothing filter (n), reconstruction from BW2 to BW5 with windowing of the data (o), reconstruction from BW2 to BW5 with windowing of the data and smoothing filter (p).}
	 \label{fig:NoInverseCrime3D} 
	\end{figure*}

\begin{figure*}[htbp]
	 \centering 
	 \scalebox{0.5} 
	 {\includegraphics{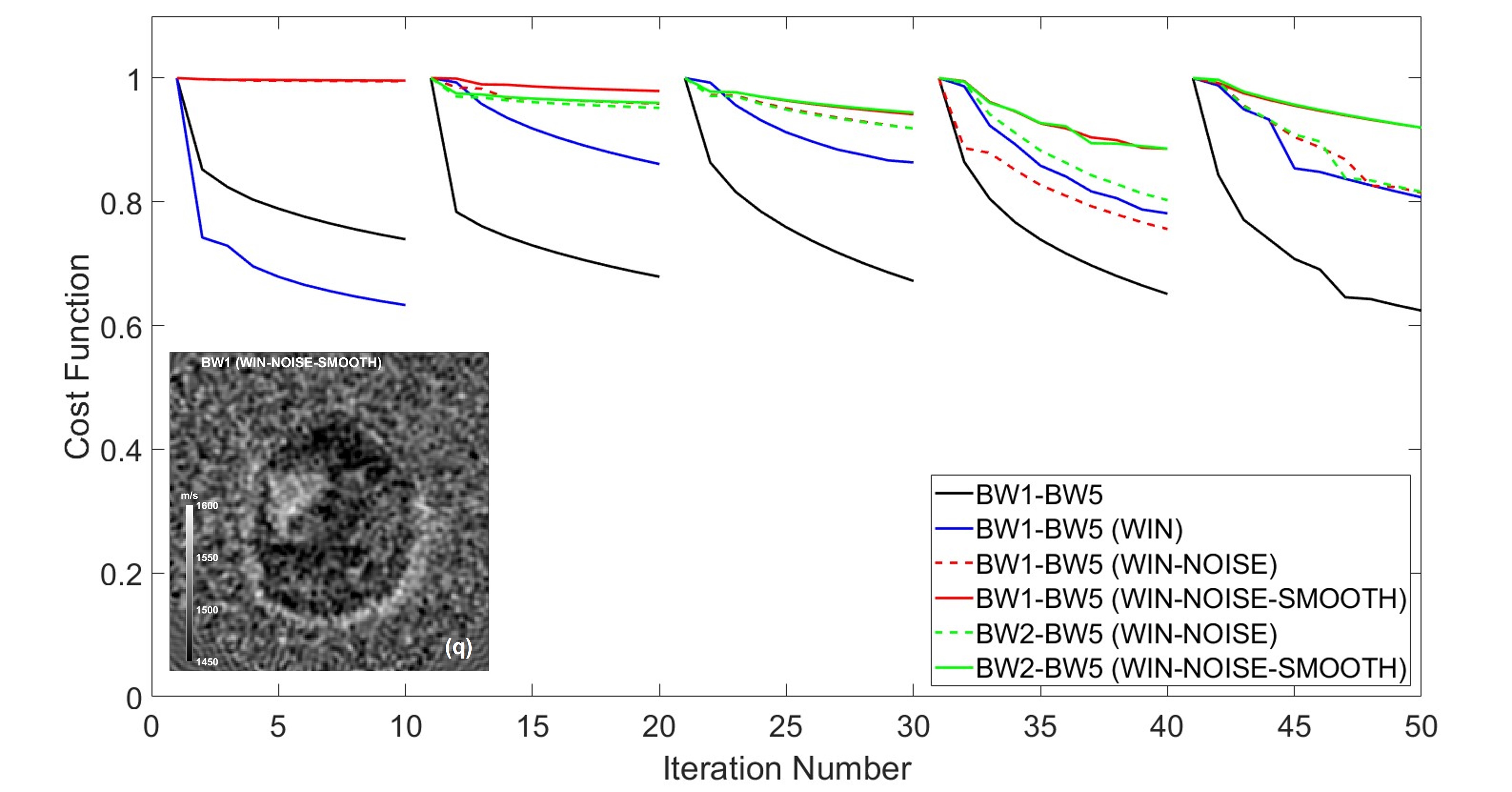}} 
	 \caption{\textit{Model mismatch  (Forward TD3D - Reconstruction FD2D). Deterministic inversion. Cost functions (noiseless and noisy data).} The image in the inset (q) shows the lack of sufficient SNR in BW1 for noisy data, which results into a very noisy image. Inversions are run for 10 iterations per bandwidth.}
	 \label{fig:NoInverseCrime3DNoise} 
	\end{figure*}

		We first invert noiseless data; reconstructed images are shown in Fig. \ref{fig:NoInverseCrime3D} (first three rows), cost functions in Fig.  \ref{fig:NoInverseCrime3DNoise}. The inversions have been run following a multi-scale, multi-frequency strategy as in the previous section (10 iterations per BW); all combinations of transmitters-receivers have initially been employed. The first notable difference between the FD2D inversion of TD2D data and the FD2D inversion of TD3D data is that the latter clearly shows circular rings that hide the true object, Fig. \ref{fig:NoInverseCrime3D} (a)-(e). These rings are evident throughout all the bandwidths, and their impact doesn't seem to disappear when inverting towards higher frequencies. To remove these rings, we consider specific combinations of transmitters-receivers, by applying a simple rectangular window (WIN) along the receivers \footnote{In array signal processing, the use of windows (along transmitters and/or receivers) is ubiquitous. In most recovery algorithms, these are used to control the side lobes: for example a Hanning window may be preferred over a rectangular window if the level of the side lobes has to be prioritized over resolution etc. These considerations are rigorously true if the imaging model (forward and reconstruction) is linear with respect to the sought-after quantity. The use of windows in the non-linear case of FWI has to be investigated numerically; in this paper, we have opted for a simple rectangular window along the selected receivers.}: as already described above, for a given transmitting element, and for the purpose of image reconstruction, we only select the receiving elements in correspondence of transmission data ($j_{Rx} = 65,...,448$ when element $i_{Tx} = 1$ and so on, Fig. \ref{fig:TxRx} (a)). This strategy effectively removes reflection data from the inversion algorithm, as well as the combinations that may contain only the direct arrival waves. In the (WIN) case, synthetic data are scaled, transmission after transmission, to match amplitude and phase at the selected receivers only. Accordingly, the corresponding images do not exhibit any ringing artifacts throughout all the bandwidths, \ref{fig:NoInverseCrime3D} (f)-(j). As the image shown in Fig. \ref{fig:NoInverseCrime3D} (j) shows a low contrast, we compress the visualization display to 1450-1600 m/s,  Fig. \ref{fig:NoInverseCrime3D} (k); this range will be used throughout all our 2D inversions of 3D synthetic data. In comparing these images with the true phantom, one has to face an extra difficulty as these 2D images are representative of the 3D volume (\emph{slice})  determined by the focusing properties of the array in elevation; comparing the final image only with the true plane co-located with the imaging plane may be a poor way of assessing the reconstruction. To overcome this difficulty, we have averaged all the 2D planes in the 3D phantom over a volume of 6 mm in height (3 mm above and below the imaging plane): although a bit naive, this allows to visualize in a 2D image the anatomy that is not located exactly in the central plane, Fig. \ref{fig:NoInverseCrime3D} (l). The latter is also displayed in the range 1450-1600 m/s. 	
	
We now consider the impact of noise, reconstructed images are shown in  Fig. \ref{fig:NoInverseCrime3D} (last row), cost functions in Fig.  \ref{fig:NoInverseCrime3DNoise}. For noisy data, we only run the reconstruction where transmitter-dependent subgroups of receiving elements are selected among all the physical ones (WIN-NOISE);  10 iterations per BW are run as in the noiseless case. Initially, we invert from BW1 to BW5 (starting from a flat initial guess, 1500 m/s). The deterioration in image quality due to noise is evident, Fig. \ref{fig:NoInverseCrime3D} (m). To cope with noise, a variety of methods can be used, for example regularization; we have opted for a mild Gaussian smoothing filter (SMOOTH) applied on the gradients at each iteration, across all bandwidths. This effectively removes some of the noise, Fig. \ref{fig:NoInverseCrime3D} (n). As the noise has been added to TD3D data by preserving the frequencies higher than 300 kHz, we expect that image quality in BW1 is poor because of the lack of SNR. This is indeed evident from the graph of the cost function in Fig. \ref{fig:NoInverseCrime3DNoise}: the reconstructed image in BW1 shows a weak and noisy shadow of the true object, Fig.  \ref{fig:NoInverseCrime3DNoise} (q). This image, when updated from BW1 to BW5, may compromise the quality of the final reconstruction BW1-BW5 (WIN-NOISE-SMOOTH), Fig.  \ref{fig:NoInverseCrime3D} (n). The reconstruction has then been re-run by inverting from BW2 to BW5, Fig. \ref{fig:NoInverseCrime3D} (o)-(p), from a  flat initial guess (1500 m/s): the final image BW2-BW5 (WIN-NOISE-SMOOTH) is the one that shows the best image quality and only minor differences with respect to the equivalent noiseless combination BW1-BW5 (WIN), Fig. \ref{fig:NoInverseCrime3D} (k). The same strategy will be used used to invert experimental data. 

			\subsubsection{Stochastic Inversion}
\begin{figure*}[htbp]
	 \centering 
	 \scalebox{0.5} 
	 {\includegraphics{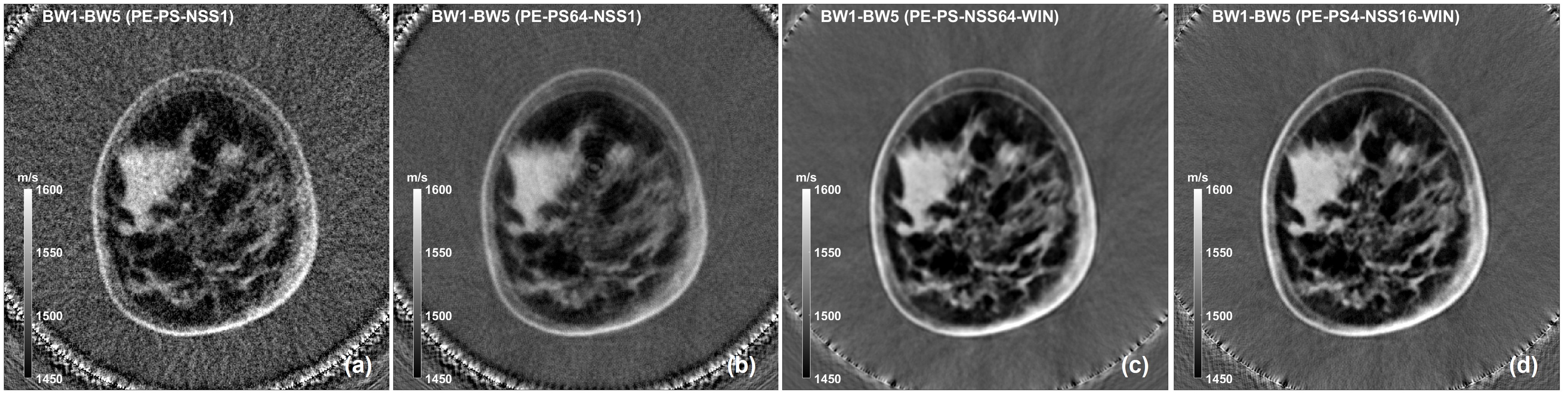}} 
	 \centering 
	 \scalebox{0.5} 
	 {\includegraphics{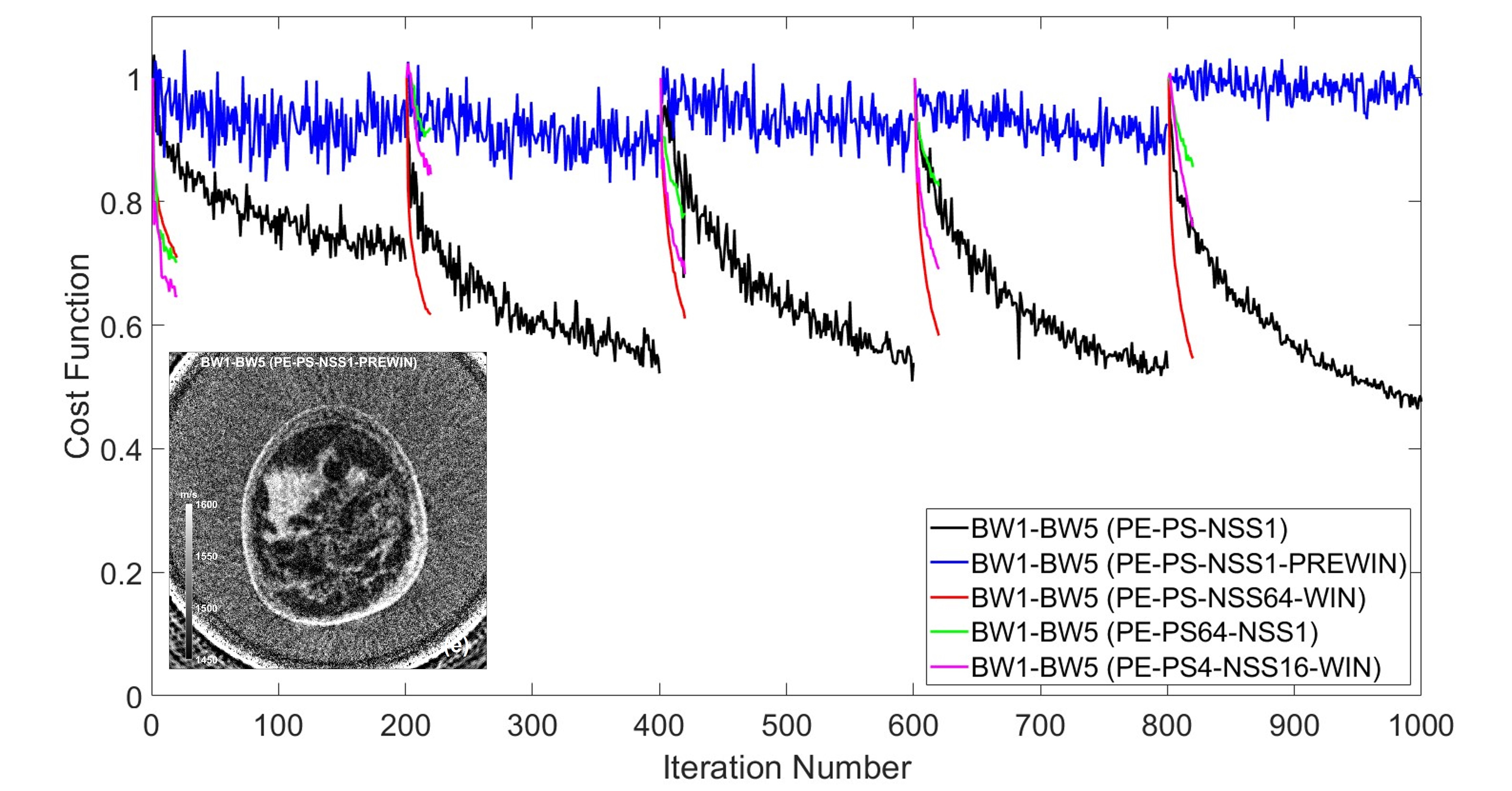}} 
	 \caption{\textit{Model mismatch  (Forward TD3D - Reconstruction FD2D). Stochastic inversion. Reconstructed images and cost functions (noiseless data).} Multi-scale, multi-frequency image reconstruction from BW1 to BW5 with all combinations of transceivers ($N_{ss} = 1$) (a), reconstruction from BW1 to BW5 with all combinations of transceivers ($N_{ss} = 1$) and stochastic ensembles ($N_{PE} = 64$) (b), reconstruction from BW1 to BW5 with multiple super-shots ($N_{ss} = 64$) and windowing of the data (c), reconstruction from BW1 to BW5 with multiple super-shots ($N_{ss} = 16$), stochastic ensembles ($N_{PE} = 4$) and windowing of the data (d). The image in the inset (e) shows the reconstruction from BW1 to BW5 with all combinations of transceivers ($N_{ss} = 1$) and \textit{pre}-windowing of the observed data: the deteriorated image quality is due to an ill-defined data fidelity term, as also evidenced by the graph of the corresponding cost function. Inversions with (without) multiple super-shots/stochastic ensembles are run for 20 (200) iterations per bandwidth.}.
	 \label{fig:NoInverseCrime3D_PE} 
	\end{figure*}

					We first invert noiseless data; images and cost functions are shown in Fig. \ref{fig:NoInverseCrime3D_PE}. All combinations of transceivers are initially employed ($N_{ss} = 1$). In this case, the texture already observed in the TD2D-FD2D case is more prominent, Fig. \ref{fig:NoInverseCrime3D_PE} (a), even after 200 iterations per BW, probably amplified by the model mismatch. With the aim to mitigate the phase encoding signal, we consider the case of one super-shot ($N_{ss} = 1$) in combination with the 				notion of stochastic ensembles ($N_{PE}= 64$), Fig. \ref{fig:NoInverseCrime3D_PE} (b); averaging the gradients 64 times is indeed effective in removing the texture. 
					
					With only one super-shot, however, some circular rings appear on top of the reconstructed object as already observed for the deterministic inversion in the case where all combinations of transceivers are used, Fig. \ref{fig:NoInverseCrime3D} (e). To further improve image quality, we then consider the case of 64 super-shots ($N_{ss} = 64, N_{PE} = 1$); in the remaining of the paper we assume that each super-shot contains $N_{Tx}^{(ss_{i})} = 86$ distinct array elements with $N_{Rx}^{(ss_{i})} = 384$ distinct array elements in the corresponding group of receivers. The central elements of the super-shots are evenly spaced along the ring-array: this means that the stochastic gradients corresponding to nearby super-shots partially overlap due to adjacent tomographic views. This effectively removes any residual texture from the image, Fig. \ref{fig:NoInverseCrime3D_PE} (c); the circular rings also disappear as in the deterministic case. Finally, we consider the case of a reduced number of super-shots ($N_{ss} = 16$) in combination with the notion of stochastic ensembles  ($N_{PE} = 4$), Fig. \ref{fig:NoInverseCrime3D_PE} (d): this combination also removes any cross-talk and any associated image texture. An immediate impact of these strategies (multiple super-shots with/without multiple stochastic ensembles) is that the number of iterations is drastically reduced, from 200 per BW to only 20 per BW: the stochastic optimization problem assumes a quasi-deterministic nature. The computational cost of the three combinations, $N_{ss} = 1$ with $N_{PE} = 64$, $N_{ss} = 64$ with $N_{PE} = 1$, $N_{ss} = 16$ with $N_{PE} = 4$, is the same as we have kept fixed the product $N_{ss} \times N_{PE} = 64$; however, image quality for the single super-shot scenario with  $N_{PE} = 64$ stochastic ensembles is inferior to the other two, as also evidenced from the graph of the cost function. 
	 \begin{figure*}[htbp]
	  \centering 
	 \scalebox{0.5} 
	 {\includegraphics{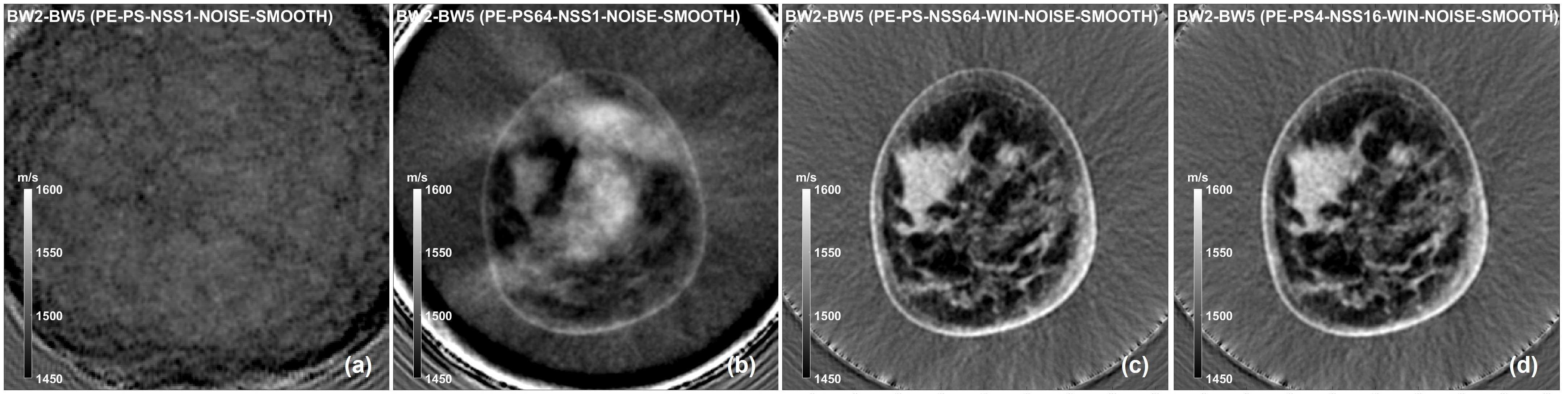}} 
	 \centering 
	 \scalebox{0.5} 
	 {\includegraphics{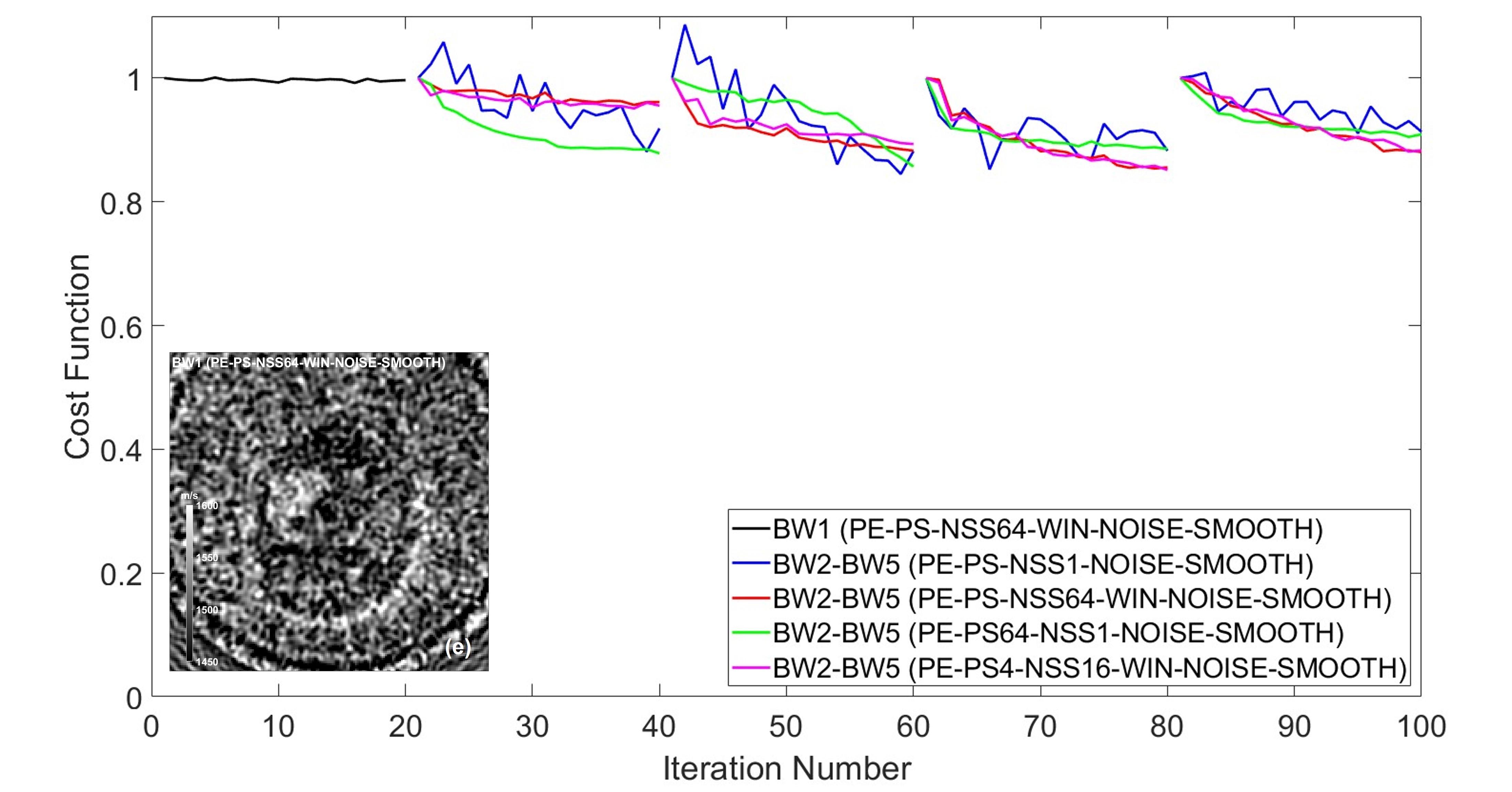}} 
	 \caption{\textit{Model mismatch  (Forward TD3D - Reconstruction FD2D). Stochastic inversion. Reconstructed images and cost functions (noisy data).} Multi-scale, multi-frequency image reconstruction from BW2 to BW5 with all combinations of transceivers ($N_{ss} = 1$) and smoothing filter (a), reconstruction from BW2 to BW5 with all combinations of transceivers ($N_{ss} = 1$), stochastic ensembles ($N_{PE} = 64$) and smoothing filter (b), reconstruction from BW2 to BW5 with multiple super-shots ($N_{ss} = 64$), windowing of the data and smoothing filter (c), reconstruction from BW2 to BW5 with multiple super-shots ($N_{ss} = 16$), stochastic ensembles ($N_{PE} = 4$), windowing of the data and smoothing filter (d). The image in the inset (e) shows the lack of sufficient SNR in BW1 for noisy data. Inversions are run for 20 iterations per bandwidth.}
	 \label{fig:NoInverseCrime3D_PE_noise} 
	 	\end{figure*}
	 	
We now invert noisy data, reconstructed images and cost functions are shown in Fig. \ref{fig:NoInverseCrime3D_PE_noise}. Reconstructions have been run as in the noiseless case; a Gaussian smoothing filter with the same strength as the one used for the deterministic reconstruction of noisy data is applied on the (averaged) gradient at each iteration. The case of one super-shot without stochastic ensembles fails completely, Fig. \ref{fig:NoInverseCrime3D_PE_noise} (a). Interestingly, the case of one super-shot with $N_{PE} = 64$ stochastic ensembles also fails, Fig. \ref{fig:NoInverseCrime3D_PE_noise} (b): this may be due to the choice of a flat initial guess. The approach with multiple super-shots is very robust, as the two reconstructions, $N_{ss} = 64, N_{PE} = 1$, Fig. \ref{fig:NoInverseCrime3D_PE_noise} (c), and $N_{ss} = 16, N_{PE} = 4$, Fig. \ref{fig:NoInverseCrime3D_PE_noise} (d), are both successful and show comparable image quality to the deterministic inversion, Fig. \ref{fig:NoInverseCrime3D} (p); the respective cost functions also reach similar level of decay. Finally, the image in the inset, Fig. \ref{fig:NoInverseCrime3D_PE_noise} (e), confirms the lack of SNR in BW1 for a stochastic reconstruction, in analogy with the deterministic case. 

In summary, the randomness due to the phase encoding signal is tamed by employing multiple super-shots (with their corresponding group of receivers) and averaging the gradients over multiple realizations of the phase encoding vectors at each iteration (stochastic ensembles). Noise is dealt with by applying the same mild smoothing filter operation on the averaged gradient as in the deterministic inversion. The same strategy will be used to invert experimental data. 

For a discussion about the possibility of keeping fixed the phase encoding vectors across all the iterations when multiple super-hots are employed, or the possibility of reducing the number of super-shots without stochastic ensembles, the reader is referred to Appendix \ref{AppendixB}.

	\subsection{Experimental Data}
	 \begin{figure*}[htbp]
	  \centering 
	 \scalebox{0.5} 
	 {\includegraphics{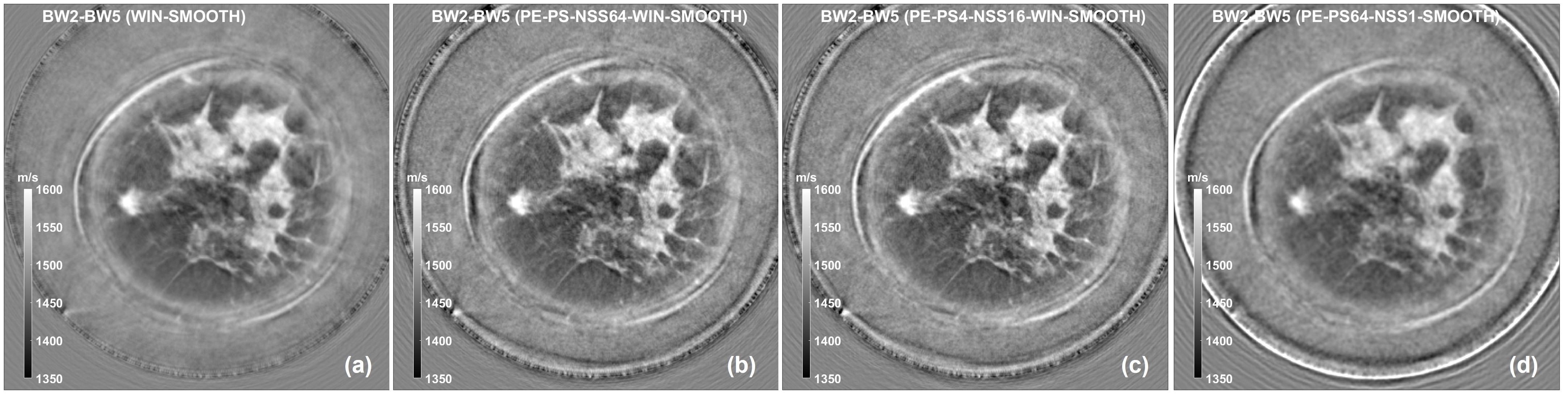}} 
	   \centering 
	 \scalebox{0.5} 
	 {\includegraphics{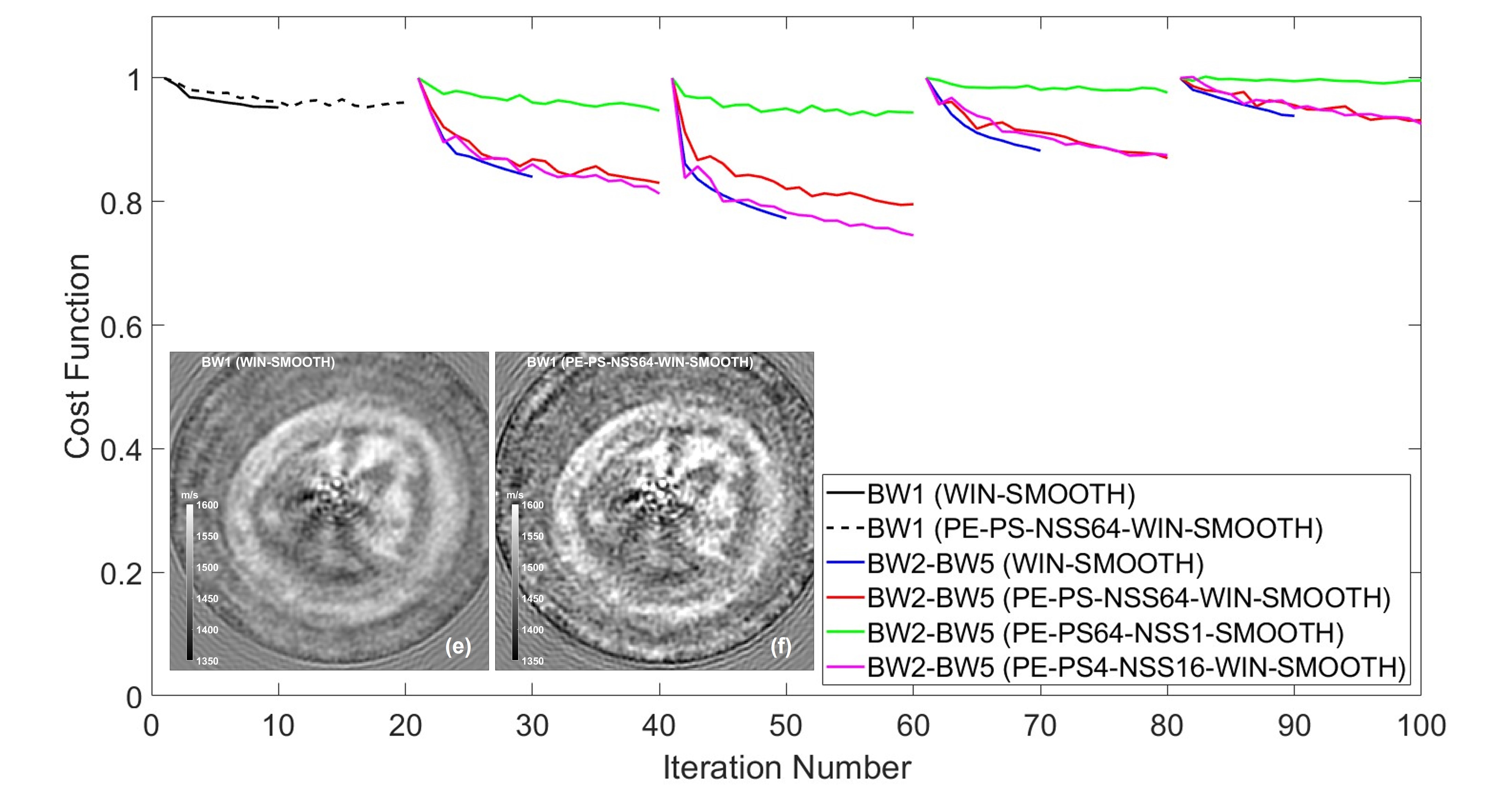}} 
	 \caption{\textit{Experimental Data, Malignancy (Reconstruction FD2D). Deterministic and stochastic inversions. Reconstructed images and cost functions.} Multi-scale, multi-frequency deterministic image reconstruction from BW2 to BW5 with windowed data and smoothing filter (a).  Multi-scale, multi-frequency stochastic image reconstruction from BW2 to BW5 with multiple super-shots ($N_{ss} = 64$), windowing of the data and smoothing filter (b), reconstruction from BW2 to BW5 with multiple super-shots ($N_{ss} = 16$), stochastic ensembles ($N_{PE} = 4$), windowing of the data and smoothing filter (c), reconstruction from BW2 to BW5 with all combinations of transceivers ($N_{ss} = 1$), stochastic ensembles ($N_{PE} = 64$) and smoothing filter (d). The images in the inset show the lack of sufficient SNR in BW1 for both reconstructions, deterministic (e) and stochastic (f). Deterministic (stochastic) inversions are run for 10 (20) iterations per bandwidth.}
	 \label{fig:Malignant1} 
	\end{figure*}
	 \begin{figure*}[htbp]
	  \centering 
	 \scalebox{0.5} 
	 {\includegraphics{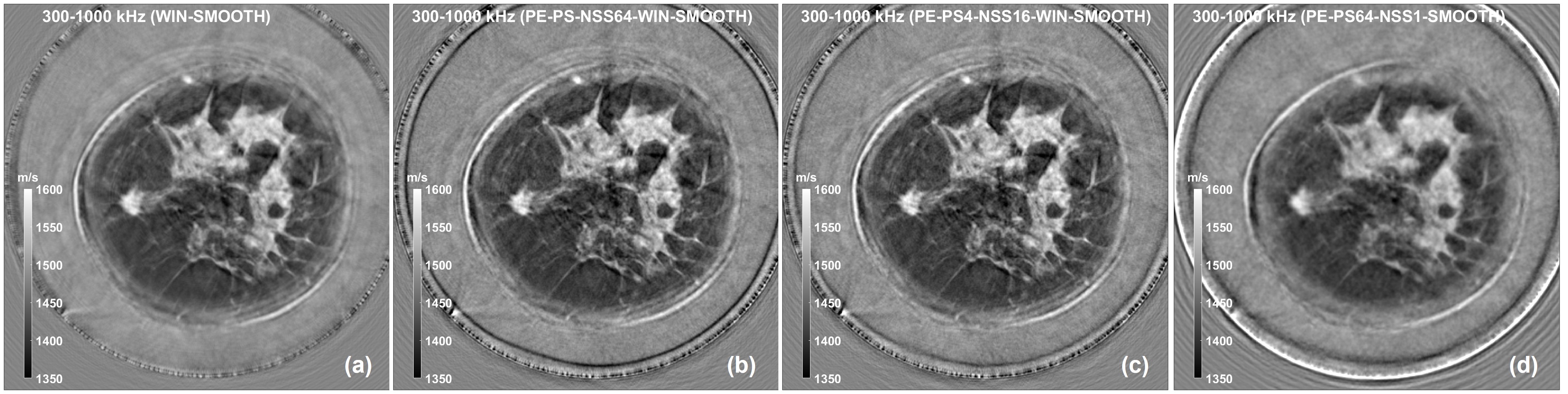}} 
	   \centering 
	 \scalebox{0.5} 
	 {\includegraphics{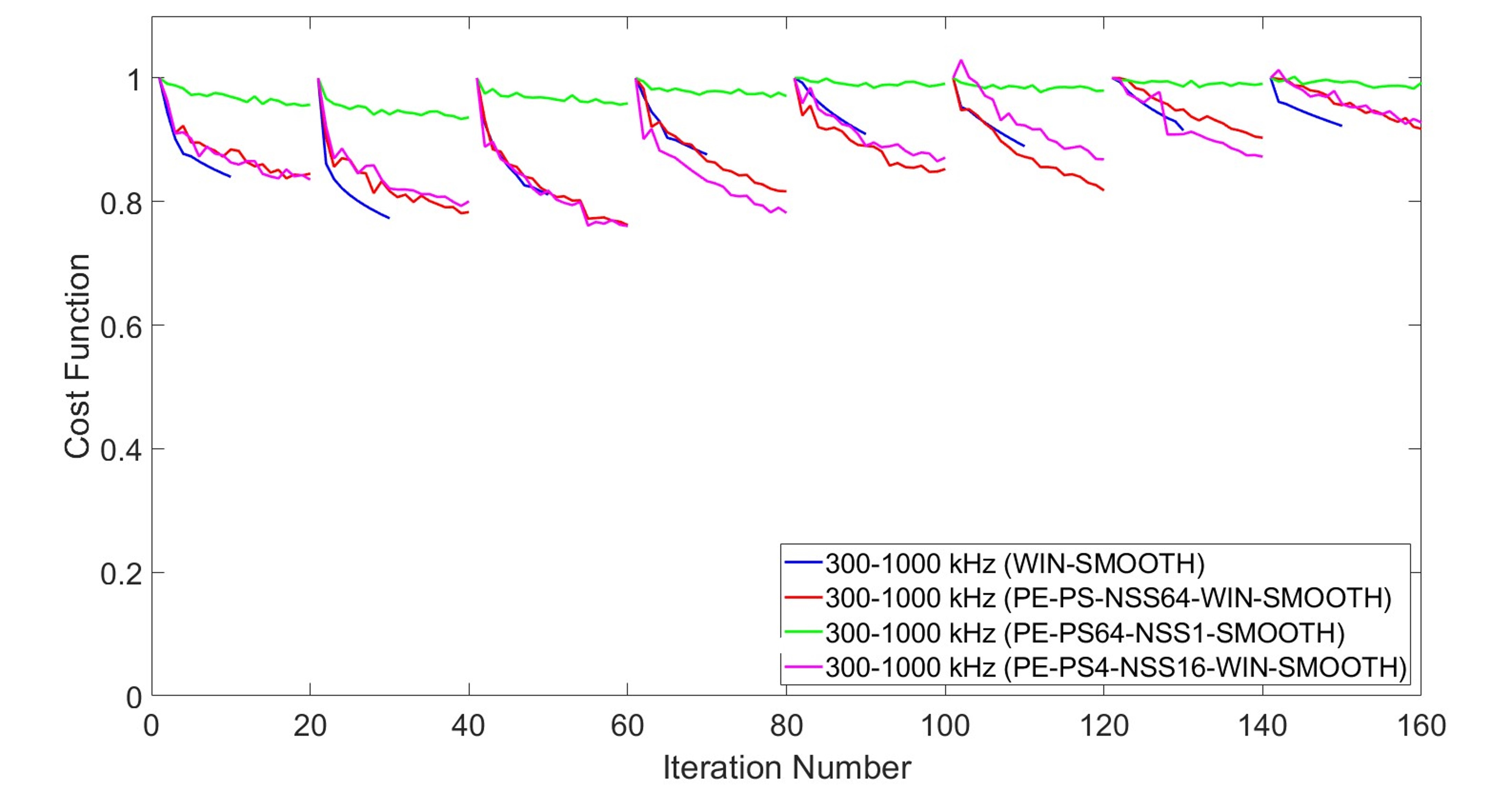}} 
 \caption{\textit{Experimental Data, Malignancy (Reconstruction FD2D). Deterministic and stochastic inversions. Reconstructed images and cost functions.} Multi-scale, \textit{single-frequency} deterministic image reconstruction from 300 kHz to 1 MHz with windowed data and smoothing filter (a).  Multi-scale, \textit{single-frequency} stochastic image reconstruction from 300 kHz to 1 MHz with multiple super-shots ($N_{ss} = 64$), windowing of the data and smoothing filter (b), reconstruction from 300 kHz to 1 MHz with multiple super-shots ($N_{ss} = 16$), stochastic ensembles ($N_{PE} = 4$), windowing of the data and smoothing filter (c), reconstruction from 300 kHz to 1 MHz with all combinations of transceivers ($N_{ss} = 1$), stochastic ensembles ($N_{PE} = 64$) and smoothing filter (d). Deterministic (stochastic) inversions are run for 10 (20) iterations per \textit{single-frequency}.}
	 \label{fig:Malignant2} 
	\end{figure*}
	 \begin{figure*}[htbp]
	   \centering 
	 \scalebox{0.5} 
	 {\includegraphics{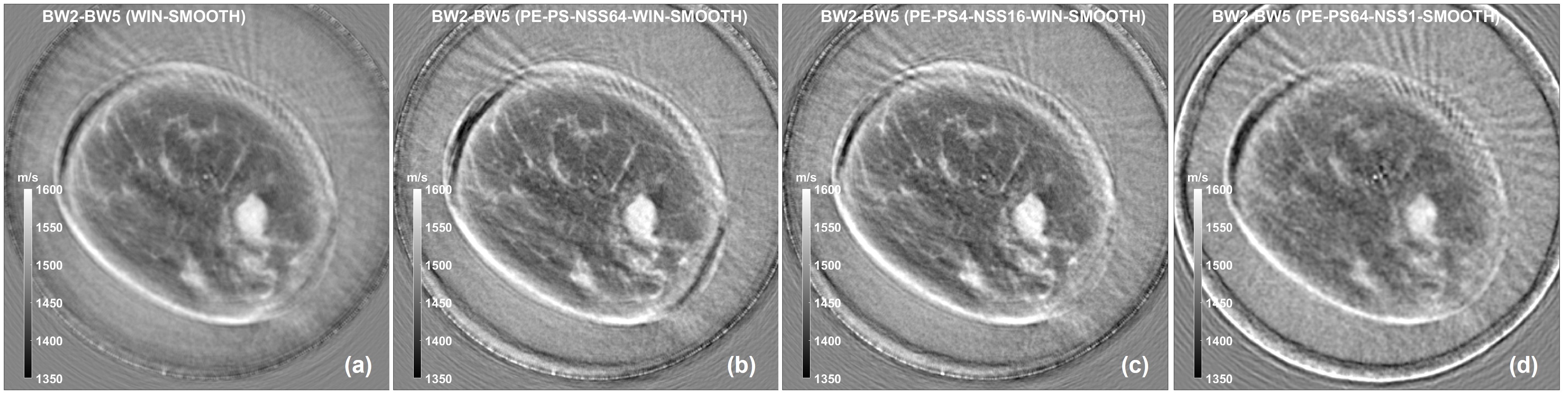}} 
	   \centering 
	 \scalebox{0.5} 
	 {\includegraphics{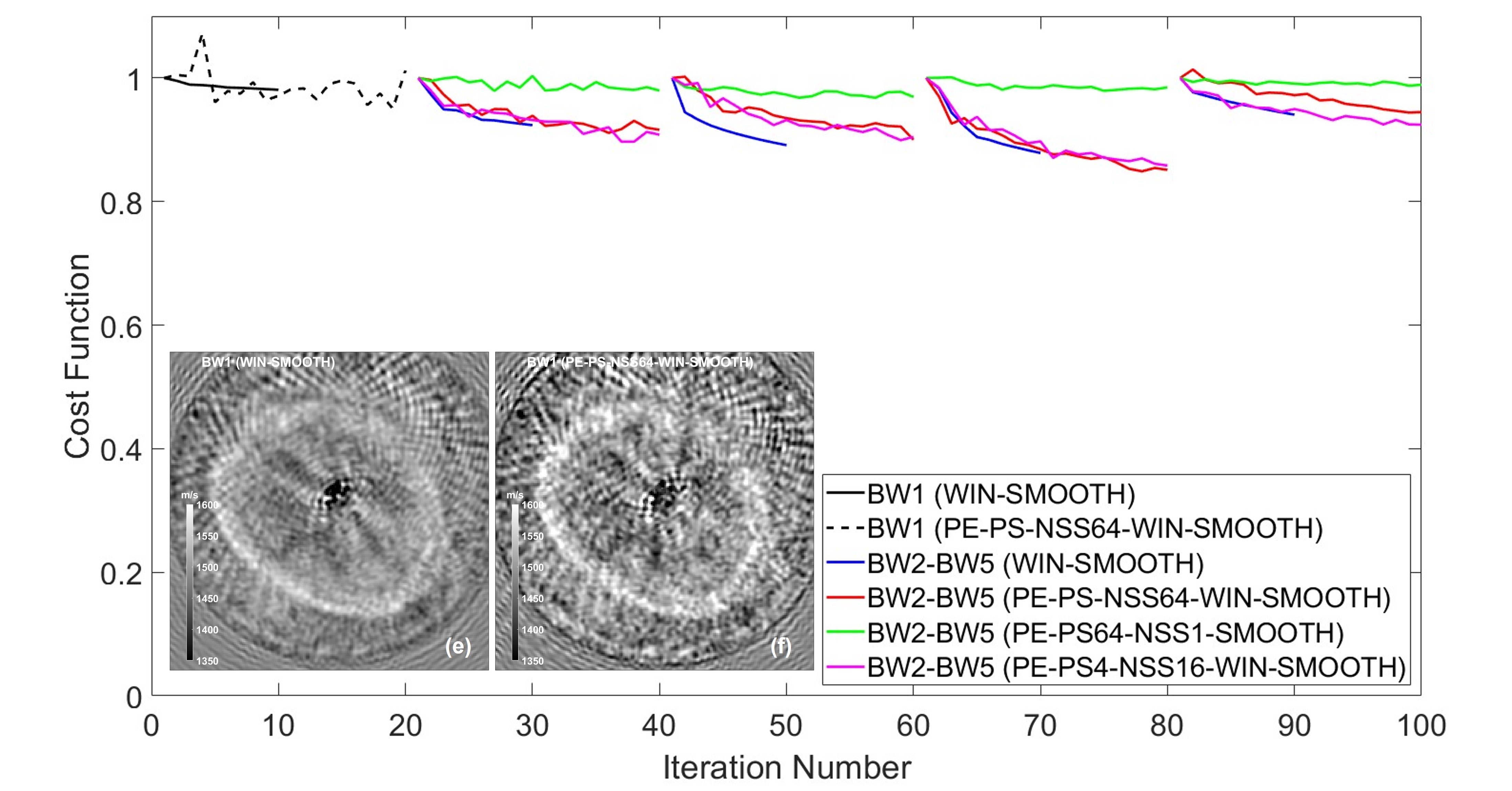}} 
	 \caption{\textit{Experimental Data, Cyst (Reconstruction FD2D). Deterministic and stochastic inversions. Reconstructed images and cost functions.} Multi-scale, multi-frequency deterministic image reconstruction from BW2 to BW5 with windowed data and smoothing filter (a).  Multi-scale, multi-frequency stochastic image reconstruction from BW2 to BW5 with multiple super-shots ($N_{ss} = 64$), windowing of the data and smoothing filter (b), reconstruction from BW2 to BW5 with multiple super-shots ($N_{ss} = 16$), stochastic ensembles ($N_{PE} = 4$), windowing of the data and smoothing filter (c), reconstruction from BW2 to BW5 with all combinations of transceivers ($N_{ss} = 1$), stochastic ensembles ($N_{PE} = 64$) and smoothing filter (d). The images in the inset show the lack of sufficient SNR in BW1 for both reconstructions, deterministic (e) and stochastic (f). Deterministic (stochastic) inversions are run for 10 (20) iterations per bandwidth.}
	 \label{fig:Cyst1} 
	\end{figure*}
		 \begin{figure*}[htbp]
	   \centering 
	 \scalebox{0.5} 
	 {\includegraphics{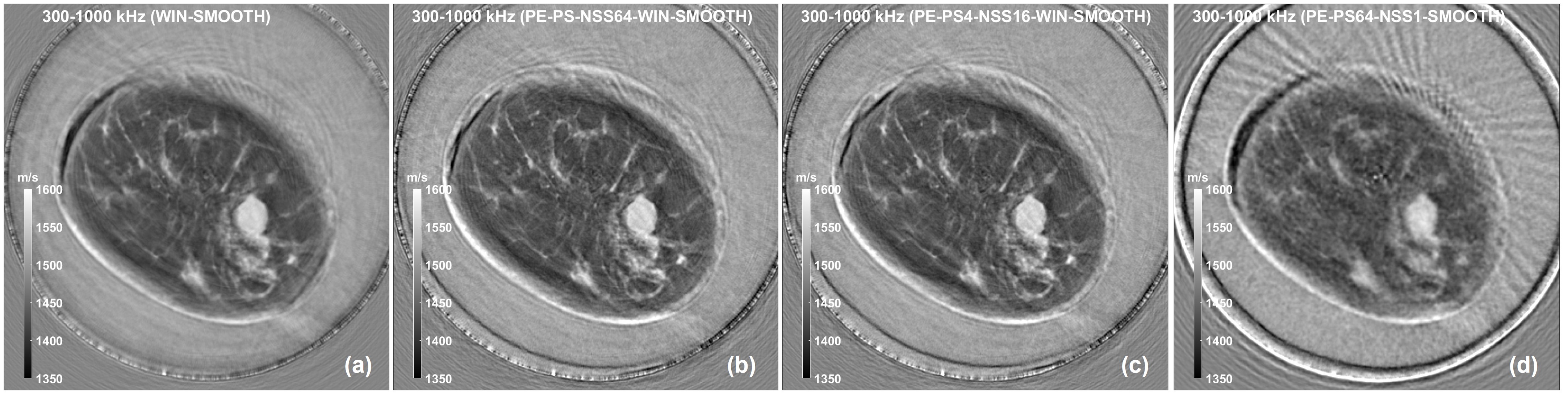}} 
	   \centering 
	 \scalebox{0.5} 
	 {\includegraphics{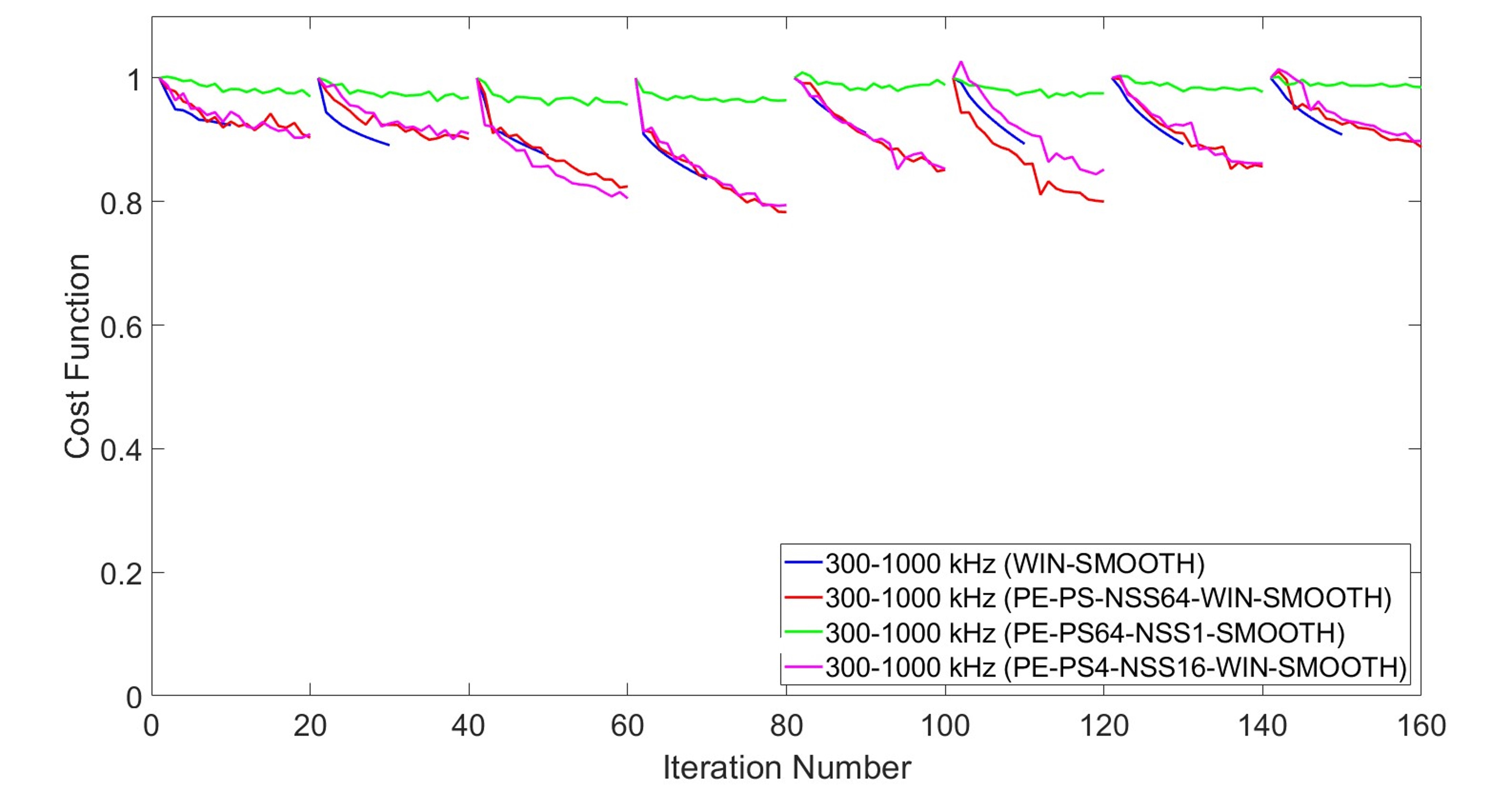}} 
 \caption{\textit{Experimental Data, Cyst (Reconstruction FD2D). Deterministic and stochastic inversions. Reconstructed images and cost functions.} Multi-scale, \textit{single-frequency} deterministic image reconstruction from 300 kHz to 1 MHz with windowed data and smoothing filter (a).  Multi-scale, \textit{single-frequency} stochastic image reconstruction from 300 kHz to 1 MHz with multiple super-shots ($N_{ss} = 64$), windowing of the data and smoothing filter (b), reconstruction from 300 kHz to 1 MHz with multiple super-shots ($N_{ss} = 16$), stochastic ensembles ($N_{PE} = 4$), windowing of the data and smoothing filter (c), reconstruction from 300 kHz to 1 MHz with all combinations of transceivers ($N_{ss} = 1$), stochastic ensembles ($N_{PE} = 64$) and smoothing filter (d). Deterministic (stochastic) inversions are run for 10 (20) iterations per \textit{single-frequency}.}
	 \label{fig:Cyst2}
	\end{figure*}
	Detailed description of the experimental data, including recommended data quality pre-processing, is given in \cite{DuricLast}. The raw dataset is a time-domain real variable of size 2112 x 512 x 512; time series are sampled at 12 MHz. Frequency domain samples are obtained with a standard FFT after zero-padding the raw data from 2112 samples to 2400 samples. This operation creates a frequency bin of 5 kHz and allows to extract the discrete frequencies from 100 kHz to 1000 kHz at the corresponding interpolated bins. The frequency domain dataset has a size of 10 x 512 x 512 complex samples. 
Deterministic and stochastic inversions are initially run exactly as in the previous section (TD3D-FD2D). The initial velocity model is a homogeneous map (1480 m/s), known because the temperature of the water bath is monitored. The Cartesian coordinates of the array elements are also known (these can be measured during the fabrication phase of the array). Results are shown in Fig. \ref{fig:Malignant1} for a malignant mass and in Fig. \ref{fig:Cyst1} for a cyst.
For both inversions, the reconstruction in BW1 shows poor image quality due to lack of sufficient SNR in BW1, Fig. \ref{fig:Malignant1} (e)(f) and  Fig. \ref{fig:Cyst1} (e)(f). The deterministic  reconstructions with windowed data, Fig. \ref{fig:Malignant1} (a) and  Fig. \ref{fig:Cyst1} (a), show comparable image quality to the stochastic reconstructions with multiple super shots and multiple stochastic ensembles ($N_{ss} = 64, N_{PE}=1$ and $N_{ss} = 16, N_{PE}=4$ respectively), Fig. \ref{fig:Malignant1} (b)(c) and  Fig. \ref{fig:Cyst1} (b)(c). The stochastic reconstruction with one super shot and multiple stochastic ensembles ($N_{ss} = 1, N_{PE}=64$), Fig. \ref{fig:Malignant1} (d) and  Fig. \ref{fig:Cyst1} (d), shows inferior image quality, as also evidenced by the graph of the cost function. Finally, all reconstructions are re-run inverting sequentially at a \textit{single-frequency}, from 300 kHz to 1 MHz in steps of 100 kHz (preserving the multi-scale approach with the number of pixels and the pixel value as in Table \ref{Tab:Tab2}). This results into a higher number of speed of sound iterations, that explains improved image quality in Fig. \ref{fig:Malignant2} and Fig. \ref{fig:Cyst2}. Results are unchanged, in particular: the stochastic reconstruction with one super-shot only has inferior image quality (again confirmed by the graph of the cost function), Fig. \ref{fig:Malignant2} (d) and Fig. \ref{fig:Cyst2} (d), the two stochastic reconstructions with multiple super-shots and multiple stochastic ensembles ($N_{ss} = 64, N_{PE}=1$ and $N_{ss} = 16, N_{PE}=4$ respectively), Fig. \ref{fig:Malignant2} (b)(c) and  Fig. \ref{fig:Cyst2} (b)(c), show very similar image quality between them and when compared against the deterministic reconstruction, Fig. \ref{fig:Malignant2} (a) and  Fig. \ref{fig:Cyst2} (a). In all the aforementioned cases, the cost functions of the two stochastic reconstructions employing multiple super-shots and multiple stochastic ensembles track each other, and both track the cost function of the deterministic reconstruction. Overall, when displayed on the 1350-1600 m/s range, image quality and anatomy are comparable to the ones published in \cite{DuricLast}. 

	\section{Discussion and Conclusions} \label{sec:end}
	
	In this paper, we have meticulously shown that a stochastic reconstruction of the speed of sound with phase encoding in the frequency domain does provide an image quality comparable, and essentially equivalent, to the one of a deterministic inversion. We have discussed in detail both the deterministic inversion algorithm and the stochastic one. Rigorous numerical evidence has been provided, carefully distinguishing the inverse crime case from more realistic scenarios, and by progressively relaxing assumptions and introducing techniques to remove the inherently stochastic nature of the phase encoding approach (multiple super-shots and stochastic ensembles) and the receiver noise (smoothing filter). In particular, the randomized inversion seems to show the same robustness to cycle skipping artifacts as the deterministic inversion (Section \ref{InvCrime} in the main text) and it seems to suffer from the same cycle skipping artifacts when the deterministic version exhibits these (Appendix \ref{AppendixA}). When combined with the notion of super-shots and multiple realizations of iteration-dependent phase encoding vectors, the randomized inversion does not seem to introduce any artifacts or structures that are not present in the true object, and the number of iterations is drastically reduced. The numerical studies are further supported by the analysis of publicly available experimental data. In particular, the randomized inversion seems to preserve the irregular/spiculated character of cancer masses (Fig. \ref{fig:Malignant1} and Fig. \ref{fig:Malignant2}) and the regular/round appearance of cysts (Fig. \ref{fig:Cyst1} and Fig. \ref{fig:Cyst2}). In a screening program for example, both inversions would pass the sensitivity test (masses are clearly detected) and the specificity test (the anatomy of the two masses can be correctly classified). 
	
The results in this paper fill an important gap, as they provide robust evidence in favor of clinically relevant image quality achievable through frequency-domain stochastic image reconstruction on \emph{experimental  USCT data}, evidence which was not fully provided in previous papers \cite{Huang}, \cite{Anastasio1}, \cite{Anastasio2}, \cite{Lucka}, \cite{Tromp} with an equivalent randomized source encoding strategy in the time-domain. The focus of these papers was on the computational gain that a stochastic inversion in the time-domain has over the deterministic inversion in the time-domain; source estimation and receivers selection were not fully addressed (with the exception of \cite{Tromp}), and the results on synthetic data were derived by mostly committing inverse crime. Whereas the results of a time-domain stochastic inversion on experimental data were shown in \cite{Anastasio1}, \cite{Anastasio2} and \cite{Tromp}, the corresponding images with deterministic techniques, TD or FD, were not included. In particular, the same cyst slice seems to be the one shown in Figs. 17-18-19 in \cite{Anastasio2} (after image transposition): the latter has been reconstructed in time-domain with a polarity source encoding scheme (PE-R) with two different optimization methods. Zooming on the images, one can clearly see radial stripes similar to the spokes of a bike wheel. These are likely to be the analogous of the image texture observed above due to the phase encoding signal: neither optimization method was able to remove them. These disappear when the regularization parameter is too strong, but in this case the entire image quality deteriorates. Based on the results of this paper, it's reasonable to assume that these spokes can be removed by averaging multiple stochastic gradients per speed of sound iteration or by employing multiple super-shots. Another source of difference in image quality in   \cite{Anastasio2} may be traced back to a sub-optimal method for source estimation, which is more challenging in TD-FWI, and even more so in a stochastic framework.				
					 In a randomized inversion with a single super-shot and a circular geometry, the concept of windowing the receivers doesn't apply in a straightforward fashion, as all the elements transmit and receive at once. This is a limitation of the randomized approach that has not been stressed enough so far in the context of ring-array USCT. A naive strategy of pre-windowing the measured data introduces a mismatch in the cost function due to ill-defined residuals. One super-shot with multiple stochastic ensembles is effective in removing most of the image texture due to the cross-talk among all the physical elements, but image quality is inferior to the deterministic one when the latter employs only transmission data. An equivalent image quality between the stochastic inversion and the deterministic one may be achieved by employing multiple super-shots, potentially in combination with multiple stochastic ensembles. Blank or faulty channels, or saturated time-series, can also be removed from the analysis within this framework, in a similar way to the deterministic inversion. A limitation of the phase encoding approach is that single transmitter, on the fly source estimation is not possible: a (super-shot-dependent) source term may be estimated by matching amplitude and phase to the (super-shot-dependent) encoded observed dataset. Nevertheless, image quality for experimental data doesn't seem to be affected.

With regard to the optimization strategy, when comparing deterministic and stochastic image quality, we have opted for keeping fixed the number of iterations per frequency bandwidth (number that it has been determined \emph{a posteriori}). This may not be a robust strategy, especially in the presence of noise: it's widely known that for iterative methods image quality strongly depends on the number of iterations (and the latter depends on the complexity of the object), and USCT makes no exception. For a deterministic inversion, one can adopt a more robust strategy where the iterations in each bandwidth are stopped when the cost function reaches a desired minimization level, with the additional rule that a maximum number of iterations (including line search ones) should be allowed. At the time of writing, the two most promising USCT systems, \cite{DuricPMB} and \cite{Wiskin2}, run a fixed number of iterations per frequency. For a randomized inversion, the stopping criterion for the cost function may be more difficult as, by definition of stochastic optimization, the cost function is allowed to increase and decrease with the iterations. The use of super-shots and stochastic ensembles, however, makes the behavior of the cost function quasi-deterministic, when compared to the variability of the cost function with one super-shot only and no stochastic ensembles. The choice of grouping a different number of frequencies per bandwidth is not necessarily robust either. In the example above, 10 iterations run in BW3 result into 10 updates of the speed of sound, Fig. \ref{fig:Malignant1} and Fig. \ref{fig:Cyst1}, whereas 10 iterations run at each of the three discrete frequencies in BW3 result in 30 updates of the speed of sound, Fig. \ref{fig:Malignant2} and  Fig. \ref{fig:Cyst2}. A more robust strategy may be to group a fixed number of frequencies per bandwidth, with the rule that the maximum frequency cannot have less than a minimum number of points per wavelength (wave-solver dependent), compatibly with memory requirements and computing times. The grouping of the frequencies proposed in this paper has intentionally been chosen sub-optimal with the aim to show the flexibility of both waveform inversion methods in terms of the number of inverted frequencies per bandwidth and corresponding points per wavelength. In particular, BW1 has been designed to contain frequencies below 300 kHz to show the lack of SNR for synthetic and experimental data and demonstrate the potential appearance of cycle skipping artifacts below 300 kHz.	

In terms of computing resources, the number of operations needed to compute the gradient are summarized in Table \ref{Tab:TabX}  \footnote{The number of operations needed for one update of the speed of sound may be higher. If line search strategies are implemented, every time the cost function is re-evaluated the computing cost will have to include all the operations needed to compute the forward field. If other strategies are implemented, for example re-calculation of the gradient or second-order methods, then the cost of the adjoint wavefield has to be included as many times as required by the optimization method.}. When the Helmholtz equation is solved via a conventional LU decomposition (the approach employed in this paper), only 1 LU decomposition is needed: this is used both to compute the forward wave-field and the adjoint wave-field. In the deterministic case, with $N_{Tx}$ physically distinct transmitting elements,  $N_{Tx}$ forward/backward substitutions are needed to compute the forward wave-field and $N_{Tx}$ forward/backward substitutions are needed to compute the adjoint wave-field. In the stochastic case, $N_{ss}$ forward/backward substitutions are needed to compute the forward wave-field and $N_{ss}$ forward/backward substitution are needed to compute the adjoint wave-field. If multiple stochastic ensembles of the phase encoding terms are used, $N_{ss}  \times N_{PE}$ forward/backward substitutions are needed for the forward wave-field and  $N_{ss}  \times N_{PE}$ forward/backward substitutions are needed for the adjoint wave-field. This results into a gain of $N_{Tx}/(N_{ss}  \times N_{PE})$ in terms of forward/backward substitutions operations. The price to pay is that, in the stochastic case, the number of iterations is higher to achieve similar image quality, although the two combinations  ($N_{ss}=64, N_{PE}=1$) and ($N_{ss}=16, N_{PE}=4$) drastically reduce the number of iterations to 20 per single frequency bandwidth. With the number of operations listed in Table \ref{Tab:TabX}, and the number of iterations as in Figs. \ref{fig:Malignant1}- \ref{fig:Cyst2}, the stochastic inversion exhibits a reduction of about 60\% in terms of computing time. With the LU decomposition and the forward/backward substitutions both run on a single GPU card (NVIDIA RTX A4000 16 GB) and the rest of the imaging algorithm implemented on the CPU in MATLAB, the stochastic inversions in Figs. \ref{fig:Malignant2}, \ref{fig:Cyst2} take around 50 minutes, against two hours for the deterministic inversions (with gradient descent with inexact line search). At the time of writing, this may be the fastest USCT imaging algorithm solving the full waveform inversion problem in 2D for the geometry of a ring-array.
	\begin{table*}
					\centering 
		\caption{Operations required to compute the gradient for deterministic and stochastic inversion.}
		\label{Tab:TabX}
		\setlength{\tabcolsep}{3pt}
		 \begin{tabular}{|p{130pt}|p{70pt}|p{60pt}|p{110pt}|}
		\hline
		Forward Wave-field  & Deterministic & Stochastic & Stochastic with Ensembles\\
		\hline
	      Helmholtz equation (LU) & &   & \\
	      	LU & 1 & 1   & 1 \\
     	      forward/backward &  $N_{Tx}$  & $N_{ss}$ &  $N_{ss} \times N_{PE}$ \\
     	      	\hline
	      Helmholtz equation  & $N_{Tx}$ &  $N_{ss}$  & $N_{ss} \times N_{PE}$ \\
	          (iterative solver)  & &  \, & \\
		\hline
		\hline
		Adjoint Wave-field  & Deterministic & Stochastic & Stochastic with Ensembles\\
		\hline
	      Helmholtz equation (LU) &   &   & \\
      	      LU & 0 & 0 & 0 \\
      	      forward/backward &  $N_{Tx}$   & $N_{ss}$ & $N_{ss}  \times N_{PE}$ \\
	      	\hline
	      Helmholtz equation & $N_{Tx}$ &  $N_{ss}$ &  $N_{ss} \times  N_{PE}$ \\
	      	          (iterative solver)  & &  \, & \\
		\hline
		\end{tabular}
		\end{table*}

When the Helmholtz equation is solved via iterative methods, the LU argument breaks downs and the cost of solving the Helmholtz equation scales linearly with the number of transmitters as in the time-domain case. In the deterministic case, then, with $N_{Tx}$ physically distinct transmitting elements, the Helmholtz equation has to be solved $2 \times N_{Tx}$ times to compute the forward and adjoint wave-fields. In the stochastic case, $2 \times N_{ss}  \times N_{PE}$ runs of the Helmholtz equations are needed. Iterative solvers of the Helmholtz equations are notoriously difficult. In the last two decades, however, there's been significant progress in developing iterative solvers based on ad-hoc preconditioners, both in 2D \cite{Erlangga}  \cite{Greenhalgh}, and in 3D \cite{Plessix3D} \cite{Operto3D}. In these papers, the authors have demonstrated the viability of iterative solvers both for the forward problem and the inverse problem, although in a deterministic setting. Because of the linearity of the wave equation in any number of space dimensions and the validity of equation (\ref{eq:aaa}) for any number of transmitting elements, the computational gain of the phase-encoding algorithm, Table \ref{Tab:TabX}, is the same in 2D and in 3D. If $N$ is the number of discretization points in each spatial coordinate and $d$ is the number of space dimensions, the acoustic impedance matrix $\boldsymbol{A}$ has size $N^{d} \times N^{d}$; this is a very large and sparse matrix. Whereas in 2D a direct solver can be applied, in 3D the complexity of the direct solver is $O(N^{6})$ and the storage required for $\boldsymbol{L}$ and $\boldsymbol{U}$ is $O(N^{5})$  \cite{Erlangga}; this makes the LU approach not feasible in 3D. Iterative solvers are alternative solutions: a matrix-vector multiplication has a complexity of $O(N^{d})$, and the overall complexity is $O(N_{it}\, N_{RHS}\, N^{d})$, where $N_{it}$ is the number of iterations and $N_{RHS}$ is the number of right hand sides (i.e. number of independent transmissions).
The images shown by QT Imaging prove that, when solving the paraxial approximation of the Helmholtz equation in 3D, a deterministic FWI algorithm is capable of removing the out of plane scattering artifacts typical of a 2D inversion; this proves the necessity for a 3D inversion. In this circumstance, and when using iterative solvers for the Helmholtz equation, a stochastic inversion has a significant computational gain over the deterministic one as $N_{RHS} = N_{ss}  \times N_{PE}$ (the convergence properties, i.e. the number of speed of sound iterations, are expected to be wave solver independent). In a full 3D reconstruction, the numbers of super-shots and stochastic ensembles may be lower, as the model-mismatch is less severe with respect to a 2D inversion, and their numbers may depend on the geometry and the noise floor of the system: a system designed to have a high SNR at the working frequencies may result into a lower values for $N_{ss}$ and $N_{PE}$. These considerations prove, once more, that the architecture of the device, the SNR at the working frequencies and the imaging algorithms have to be designed together. The feasibility of iterative solvers for the Helmholtz equation in 3D for USCT applications, however, has to be proven, especially in terms of computing times. 

The investigation of the previous arguments will be the subject of future publications.
		        \section{Personal Statement} 
		        This manuscript has been sent for review on the 21st of March 2025, and in its current/revised form on the 30th of May 2025. 
		\appendix
		        \section{A counter-example with cycle skipping (phase wrapping)}  \label{AppendixA}
		         \begin{figure*}[htbp]
		           \centering 
	 \scalebox{0.5} 
	 {\includegraphics{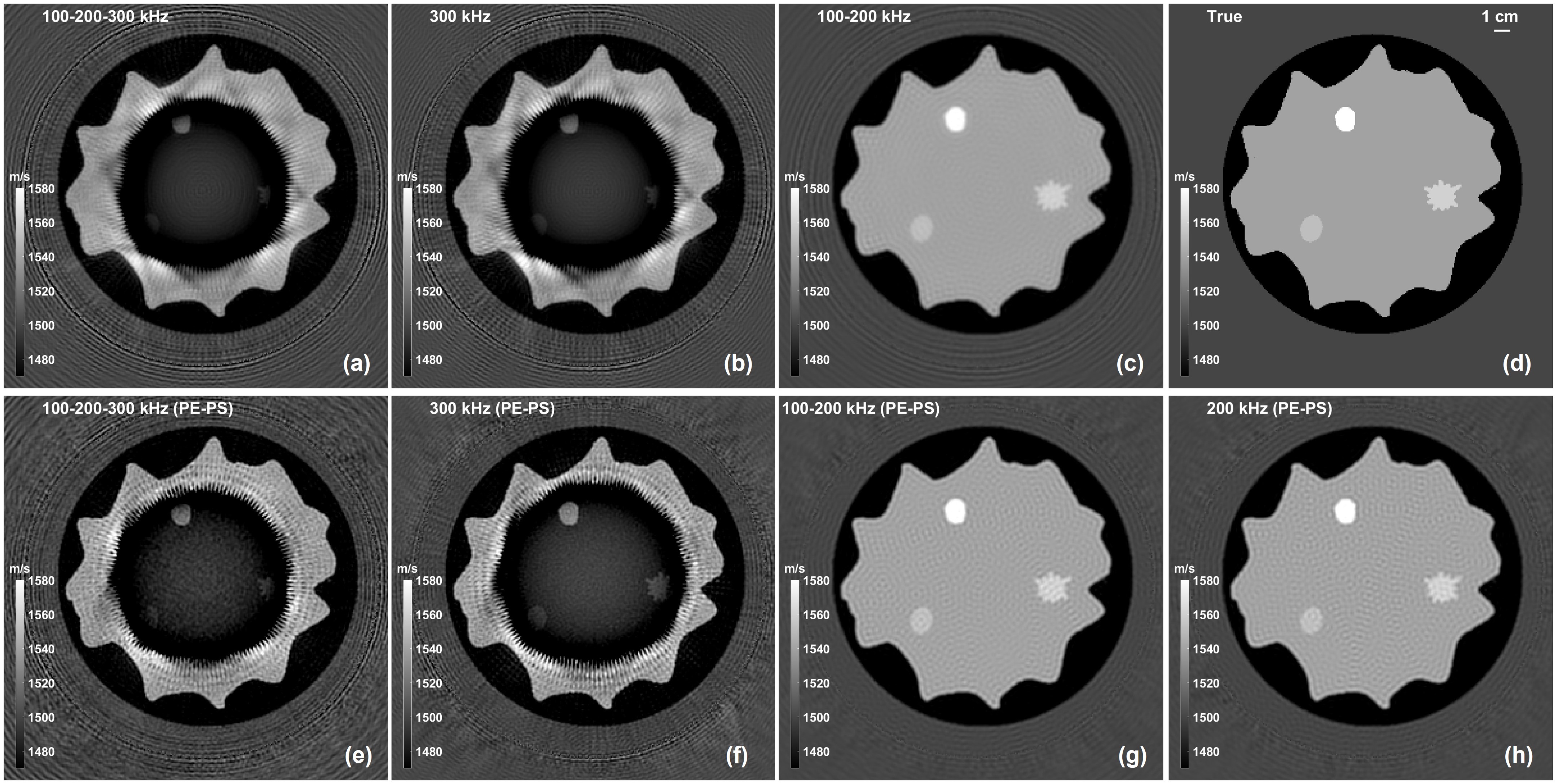}} 
	 \caption{\textit{Inverse Crime with phase wrapping (Forward FD2D - Reconstruction FD2D). Deterministic and stochastic inversions. Reconstructed images.} Reconstruction fails when inverting simultaneously at 100-200-300 kHz, or at a single frequency of 300 kHz, deterministic (a) and (b), stochastic (e) and (f), but it is successful when inverting simultaneously at 100-200 kHz, deterministic (c), stochastic (g), or at a single frequency of 200 kHz, stochastic (h).}
	 \label{fig:InverseCrimeAppendix}
	\end{figure*}
		        In this Appendix, we provide a counterexample to the discussion in the main text, Section \ref{InvCrime}. We consider a different numerical phantom, inspired by the experimental phantom described in \cite{DuricCS} and \cite{Anastasio1}; for simplicity we limit our discussion to the full inverse crime case (FD2D-FD2D). Data generation and image formation follow the same procedure as in Section \ref{InvCrime}. Results are shown in Fig. \ref{fig:InverseCrimeAppendix}. The deterministic inversion fails inverting three frequencies simultaneously (100-200-300 kHz), Fig. \ref{fig:InverseCrimeAppendix} (a), it fails inverting a single frequency at 300 kHz, Fig. \ref{fig:InverseCrimeAppendix} (b), but it is successful inverting at 100-200 kHz, Fig. \ref{fig:InverseCrimeAppendix} (c): this suggests that the numerical problem becomes non-convex for frequencies higher than 300 kHz (for a given geometry and the tested range of speed of sound values 1450 - 1580 m/s). The stochastic inversion with phase encoding exhibits similar behavior, Fig. \ref{fig:InverseCrimeAppendix} (e)-(h): this proves that a randomized strategy preserves, in a stochastic sense, the convexity property for the same combination of frequencies, geometry and range of values for the underlying speed of sound. In other words, the frequency threshold in order to avoid the cycle-skipping (phase wrapping) problem is preserved with a randomized inversion. This completes the analysis in the main text.

	        \section{Iteration-independent phase encoding, super-shots and stochastic ensembles} \label{AppendixB}
	         \begin{figure*}[htbp]
    \centering 
	 \scalebox{0.5} 
	 {\includegraphics{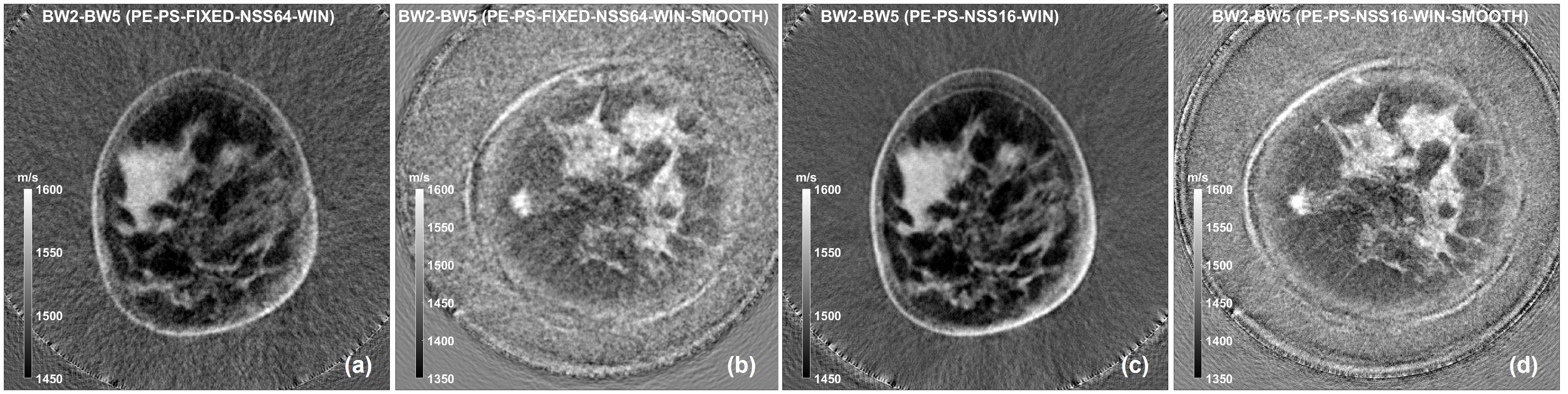}} 
	 \caption{\textit{Stochastic reconstruction with multiple super-shots and fixed phase-encoding vectors. Stochastic reconstruction with multiple super-shots without stochastic ensembles. Reconstructed images for noiseless data (Forward TD3D) and experimental data (Malignancy).} Multi-scale, multi-frequency stochastic image reconstruction from BW2 to BW5 with multiple super-shots ($N_{ss} = 64$), windowing of the data and fixed encoding vectors, synthetic (a), experimental (b).  Multi-scale, multi-frequency stochastic image reconstruction from BW2 to BW5 with multiple super-shots ($N_{ss} = 16$), without multiple stochastic ensembles ($N_{PE} = 1$) and with windowing of the data, synthetic (c), experimental (d). Inversions are run for 20 iterations per bandwidth.}
	 \label{fig:PEAppendix} 
	 \end{figure*}
	      In Section \ref{InvCrime}, we have shown that, under the hypothesis of a single super-shot ($N_{ss}=1$) and without stochastic ensembles ($N_{PE}=1$), the stochastic optimization fails if the encoding vectors are not re-drawn at each speed of sound iteration. In this Appendix, we show the impact of generating the phase encoding terms only once and keeping them fixed for all iterations, in combination with the notion of multiple super-shots ($N_{ss}=64$). In this case, the optimization problem is deterministic (regardless of how the $\boldsymbol{a}$ are generated), because the phase encoding terms are kept constant for all iterations. In fact, the cost functions (not shown) exhibit a strictly monotonically decreasing behavior.  Reconstructed images are shown in Fig. \ref{fig:PEAppendix}, for noiseless TD3D synthetic data (a), and for the malignancy experimental dataset (b). The evident texture on top of the image clearly suggests that independent realizations of the phase-encoding vectors are required at each speed of sound iteration. Re-generating the phase encoding terms at each iteration in combination with the notion of multiple super-shots is very effective both in removing any cross-talk \emph{and} in reducing the number of iterations per frequency bandwidth. Similar investigations in the context of seismic imaging were considered in \cite{Leeuwen1} - \cite{Leeuwen2}, with somewhat different conclusions. The results in these two papers, however, were derived considering only the FD2D-FD2D inverse crime scenario.
	      
	      Finally, we test the impact on image quality under the hypothesis of a limited number of super-shots ($N_{ss}=16$) \textit{without} multiple stochastic ensembles  ($N_{PE}=1$). This scenario is desirable, because few Helmholtz equations are solved per speed of sound iteration. However, both the synthetic image, Fig. \ref{fig:PEAppendix} (c), and the experimental one, Fig. \ref{fig:PEAppendix} (d), are again dominated by an evident texture due to the phase-encoding signal. A trade-off between number of super-shots and number of stochastic ensembles is needed; the two choices made in the main text of $N_{ss}=64,  N_{PE}=1$ and $N_{ss}=16,  N_{PE}=4$ seem to provide equivalent results for the same computational cost ($N_{ss} \times N_{PE}=64$). 

	\bibliography{PhaseEncodingFrequencyDomain}
	
	\end{document}